\documentclass[english]{IEEEtran}
\usepackage[T1]{fontenc}
\usepackage[latin9]{inputenc}
\usepackage{color}
\usepackage{array}
\usepackage{float}
\usepackage{url}
\usepackage{amsmath}
\usepackage{amsthm}
\usepackage{amssymb}
\usepackage{graphicx}

\makeatletter

\providecommand{\tabularnewline}{\\}
\floatstyle{ruled}
\newfloat{algorithm}{tbp}{loa}
\providecommand{\algorithmname}{Algorithm}
\floatname{algorithm}{\protect\algorithmname}

\theoremstyle{plain}
\newtheorem{thm}{\protect\theoremname}
\theoremstyle{plain}
\newtheorem{lem}[thm]{\protect\lemmaname}
\theoremstyle{plain}
\newtheorem{cor}[thm]{\protect\corollaryname}
\theoremstyle{definition}
\newtheorem{defn}[thm]{\protect\definitionname}

\usepackage{babel}
\usepackage{cite}
\usepackage{algorithm}
\usepackage{algorithmic}

\providecommand{\definitionname}{Definition}
\providecommand{\lemmaname}{Lemma}
\providecommand{\theoremname}{Theorem}

\providecommand{\corollaryname}{Corollary}

\@ifundefined{showcaptionsetup}{}{%
 \PassOptionsToPackage{caption=false}{subfig}}
\usepackage{subfig}
\makeatother

\usepackage{babel}
\providecommand{\corollaryname}{Corollary}
\providecommand{\definitionname}{Definition}
\providecommand{\lemmaname}{Lemma}
\providecommand{\theoremname}{Theorem}

\begin{document}
\title{An Information-Theoretic Characterization of MIMO-FAS: Optimization,
Diversity-Multiplexing Tradeoff and $q$-Outage Capacity\thanks{The work of W. K. New, K. K. Wong and K. F. Tong is supported by the
Engineering and Physical Sciences Research Council (EPSRC) under grant
EP/W026813/1.}\thanks{The work of C.-B. Chae is supported by the Institute for Information
and Communication Technology Promotion (IITP) grant funded by the
Ministry of Science and ICT (MSIT), Korea (No. 2021-0-02208, No. 2021-0-00486).}\thanks{The work of H. Xu is supported by the European Union's Horizon 2020
Research and Innovation Programme under Marie Sklodowska-Curie Grant
No. 101024636.}\thanks{W. K. New (email: ${\rm a.new@ucl.ac.uk}$), K. K. Wong (corresponding
author, email: ${\rm kai\text{-}kit.wong@ucl.ac.uk}$), H. Xu (email:
${\rm hao.xu@ucl.ac.uk}$), and K. F. Tong (email: ${\rm k.tong@ucl.ac.uk}$)
are with the Department of Electronic and Electrical Engineering,
University College London, London, WC1E 6BT, United Kingdom. C.-B.
Chae (email: ${\rm cbchae@yonsei.ac.kr}$) is with the School of Integrated
Technology, Yonsei University, Seoul 03722 Korea. K. K. Wong is also
affiliated with Yonsei Frontier Lab., Yonsei University, Seoul 03722,
Korea.}}
\author{Wee Kiat New, \textit{Member, IEEE}, Kai-Kit Wong, \textit{Fellow,
IEEE}, Hao Xu, \textit{Member, IEEE,}\\
 Kin-Fai Tong, \textit{Fellow, IEEE}, \textit{and} Chan-Byoung Chae,
\textit{Fellow, IEEE}}
\maketitle
\begin{abstract}
Multiple-input multiple-output (MIMO) system has been the defining
mobile communications technology in recent generations. With the ever-increasing
demands looming towards the sixth generation (6G), we are in need
of additional degrees of freedom that deliver further gains beyond
MIMO. To this goal, fluid antenna system (FAS) has emerged as a new
way to obtain spatial diversity using reconfigurable position-switchable
antennas. Considering the case with more than one ports activated
on a 2D fluid antenna surface at both ends, we take the information-theoretic
approach to study the achievable performance limits of the MIMO-FAS.
First of all, we propose a suboptimal scheme, referred to as QR MIMO-FAS,
to maximize the rate at high signal-to-noise ratio (SNR) via joint
port selection, transmit and receive beamforming and power allocation.
We then derive the optimal diversity and multiplexing tradeoff (DMT)
of MIMO-FAS. From the DMT, we highlight that MIMO-FAS outperforms
traditional MIMO antenna systems. Further, we introduce a new metric,
namely $q$-outage capacity, which can jointly consider rate and outage
probability. Through this metric, our results indicate that MIMO-FAS
surpasses traditional MIMO greatly. 
\end{abstract}


\section{Introduction}

\subsection{Background}

\IEEEPARstart{S}{ixth-generation} (6G) mobile communication seeks
to push the key performance indicators (KPIs) way beyond what the
current fifth generation (5G) promises to offer. Such upgrade will
require new technologies that can achieve more from the same amount
of bandwidth. Presently, the dominating technology has been multiple-input
multiple-output (MIMO), which also comes in the form of multiuser
MIMO and massive MIMO. In 6G, the desire is to exceed MIMO \cite{Tariq-2020,9770295}.

To achieve this very ambitious goal, one emerging idea is fluid antenna
system (FAS) \cite{wong2022bruce}. FAS represents any software-controllable
fluidic, conductive or dielectric structure that can adjust its shape
and position to reconfigure the gain, radiation pattern, operating
frequency and other radiation characteristics. This is now feasible,
thanks to the recent advances in utilizing flexible conductive materials
such as liquid metals or ionized solutions \cite{Huang-2021access},
switchable pixels \cite{Besoli-2011,Song-2014}, and stepper motors
for antennas \cite{Ismail-1991,Basbug-2017}. The concept of fluid
antenna includes all forms of movable and non-movable flexible-position
antennas.

Unlike traditional antenna that is placed at a fixed location, fluid
antenna is able to switch its location almost instantly in a limited
space. The most basic single fluid antenna consists of one radio frequency
(RF)-chain and $N$ preset locations (also known as ports) that are
distributed in a given space \cite{wong2022bruce}. The radiating
element of the fluid antenna can switch its position to obtain a higher
rate, lower outage probability, less interference and other desirable
performance gains depending on the applications. As the ports can
be placed closely to each other, the channels of these ports are strongly
correlated and thus spatial correlation plays a crucial role in FAS.

The main implementation designs for FAS are: i) liquid-based fluid
antenna and ii) RF pixel-based fluid antenna. In the liquid-based
fluid antenna, each liquid droplet can precisely switch its position
by controlling the electric field using the thin conductive lines
on top of the dielectric layer while other technology may use an electronically
controlled pump to shift the position of a fluid radiating element
in a tube. On the other hand, in the RF pixel-based fluid antenna,
the RF pixels can be turned on-and-off instantly regardless of the
surface area. One or several pixels when on, form an antenna port
for transmission or reception like a standard antenna. Besides, each
activated port is connected to an RF-chain, operating like a conventional
antenna. In short, the basic principle of FAS is to exploit the dynamic
nature of fluid antenna to achieve ultimate flexibility for diversity
and multiplexing gains.\footnote{It is worth pointing out that FAS does not necessarily use `fluid'
materials for antenna and in wireless communications that requires
adaptation in time of milliseconds or less, reconfigurable pixels
are more relevant.}

Due to its unprecedented benefits, single-user, single-input single-output
(SISO)-FAS has recently been investigated under different scenarios
and assumptions. Specifically, as the number of ports increases, \cite{9264694}
showed that the outage probability of FAS could be reduced drastically
while \cite{9131873} demonstrated that FAS could significantly improve
the ergodic capacity. Motivated by these works, \cite{9833952} derived
the level crossing rate, \cite{9715064} devised a port selection
algorithm by observing the channels of a few ports, \cite{9771633}
investigated the performance of FAS over general correlated channels
and \cite{tlebaldiyeva2023outage} analyzed the outage probability
of FAS for Terahertz communications while selection combining and
maximum gain combining were further considered. Optimistic results
were obtained in these works but it was illustrated in \cite{khammassi2022new}
that the outage probability of FAS could only be reduced to a floor
when a more accurate spatial correlation model was adopted. The recent
work \cite{new2022fluid} explained such limitations at an intuitive
level and revealed that the performance of FAS was generally determined
by the available space. Furthermore, only in certain cases might FAS
achieve a similar outage probability as compared to the classical
maximal ratio combining (MRC) system.

The unique ability of switching the antenna position finely in FAS
can also be exploited to mitigate interference, which would be impractical
in traditional antenna selection systems. Recently, \cite{9992289}
investigated orthogonal multiple access to serve multiple users with
fluid antennas while \cite{zhu2023movable} used a space division
multiple access approach to minimize the user transmit power. Nevertheless,
an arguably more interesting idea is the fluid antenna multiple access
(FAMA) scheme \cite{9650760} where the rationale is to exploit the
moment of deep fades in the spatial domain to alleviate inter-user
interference. FAMA is classified into slow FAMA and fast FAMA in which
the former switches its port when the channel changes \cite{10018377}
and the latter switches its port on a symbol-by-symbol basis \cite{9953084}.
Most recently, the outage probability for two-user FAMA was revisited
in \cite{OutageFAMA}.

In summary, FAS has shown promises but much is still not well understood.
For example, the performance of FAS itself can be lifted if more than
one ports are activated. For a point-to-point communication channel,
we refer to the system where both ends are equipped with a multi-port
FAS, as MIMO-FAS which is also known as fluid MIMO or flexible MIMO
in \cite{wong2022bruce}. \textcolor{black}{Note that multiple ports
can also be activated in a two dimensional (2D) surface using liquid-based
fluid antenna or RF pixel-based fluid antenna. The schematics of the
MIMO-FAS designs were discussed in \cite{wong2022bruce}.} Compared
to a traditional MIMO antenna selection system in which the number
of antennas is limited in a given surface (at least half wavelength
separation between the antennas) and the antennas are fixed in positions,
MIMO-FAS is distinct in the sense that the positions of the radiating
elements can be dynamically and finely adjusted and that the number
of preset locations (i.e., ports) within a given surface can be arbitrarily
large, which yields additional gains. Note that FAS has also been
proposed for multiple access recently \cite{10018377,9953084,OutageFAMA},
where the fine resolution of FAS is absolutely essential and conventional
antenna selection would be unable to cope.

It is anticipated that the capacity and reliability of MIMO-FAS will
be improved over the SISO counterpart. In fact, a related work showed
that the capacity of a movable antenna system (which can be interpreted
as MIMO-FAS with movable fluid antennas) could be improved up to $30.3\%$
as compared to traditional MIMO systems \cite{ma2022mimo}. In the
study, however, spatial correlation due to rich scattering between
the antenna positions was not considered. More importantly, the optimal
diversity and multiplexing tradeoff (DMT) of MIMO-FAS is unknown.

In information theory, the optimal DMT can be employed as a unified
framework to compare the performance of different multiple-antenna
systems \cite{1197843}. More concretely, it focuses on the asymptotic
high signal-to-noise ratio (SNR) regime and a scheme is then said
to achieve a multiplexing gain of $r$ and a diversity gain of $d\left(r\right)$
if the rate of the system scales like $r\log{\rm SNR}$ and its outage
probability decays like ${\rm SNR}^{-d\left(r\right)}$. It is known
that $r$ cannot exceed the total degrees of freedom of the channel
and $d\left(0\right)$ is limited by the maximal diversity gain, i.e.,
total number of independent channels. In between the two extremes,
a system must tradeoff each type of gains.

\subsection{Contributions}

Motivated by the above, this paper analyzes the performance of MIMO-FAS
with the goal of gaining useful insights for designing an efficient
MIMO-FAS. To this end, we first develop a system model of MIMO-FAS
while taking into account of the spatial correlation effect. \textcolor{black}{To
characterize the performance limits of MIMO-FAS, we consider a rich
scattering environment since it is well known that multipath can help
to improve the diversity and multiplexing gains.}\footnote{\textcolor{black}{Note that rich scattering can help to improve the
performance of any MIMO systems. This includes MIMO-FAS, traditional
MIMO and MIMO antenna selection. Therefore, if the number of scatterers
is small, one may further consider using a reconfigurable intelligent
surface to create artificial scatterers to improve the performance
of any MIMO systems.}} We then propose a suboptimal scheme that maximizes the rate of MIMO-FAS
through joint port selection, transmit and receive beamforming and
power allocation at high SNR. Based on this scheme, we derive the
optimal DMT of MIMO-FAS to reveal the fundamental limits of MIMO-FAS
from an information-theoretic viewpoint. From the analytical results,
we further study the effects of different MIMO-FAS parameters and
reveal the superiority of MIMO-FAS over traditional MIMO and MIMO
antenna selection in terms of DMT. In addition, we introduce a new
metric, referred to as $q$-outage capacity, to showcase the benefits
of MIMO-FAS. Our main contributions are summarized as follows: 
\begin{itemize}
\item We develop a system model for MIMO-FAS with a 2D fluid antenna surface
at both ends while taking into account of the spatial correlation
of the ports. In particular, we employ a simple yet accurate channel
model that considers the spatial correlation in a three-dimensional
(3D) scattering environment. Based on this channel model, we introduce
several system parameters such as active ports, beamforming matrices
and power allocation. The achievable rate of MIMO-FAS is then derived
where its expression resembles the rate of a traditional MIMO system. 
\item Also, we formulate a non-convex optimization problem to maximize the
rate of MIMO-FAS via joint port selection, transmit and receive beamforming
and power allocation. We show that the global optimal solution can
be obtained using an exhaustive search, singular value decomposition
(SVD) and waterfilling power allocation, at the expense of a non-polynomial
time complexity. To reduce the time complexity, we propose the QR
MIMO-FAS scheme that maximizes the rate of MIMO-FAS at high SNR via
suboptimal port selection, beamforming and power allocation. It is
shown that QR MIMO-FAS has a polynomial time complexity. 
\item Furthermore, we derive the outer bound of the DMT of MIMO-FAS. By
using the outer bound and QR MIMO-FAS, we obtain the optimal DMT of
MIMO-FAS. In this process, we prove that the spatial correlation matrix
$\boldsymbol{J}_{s}$ can be represented by a finite-size matrix $\boldsymbol{J}_{{\rm red}}^{s}$
even if the number of ports increases to infinity. Afterwards, we
propose methods to estimate the size of $\boldsymbol{J}_{{\rm red}}^{s}$
and linearly transform between $\boldsymbol{J}_{s}$ and $\boldsymbol{J}_{{\rm red}}^{s}$
with proof of certificates. 
\item Extensive results are provided to highlight the effects of several
MIMO-FAS parameters. In the discussions, we provide useful insights
for designing an efficient MIMO-FAS. Although MIMO-FAS provides rate
improvements over the traditional MIMO antenna systems, we highlight
that the superiority of MIMO-FAS actually lies in the diversity gain.
Specifically, the diversity gain of MIMO-FAS for a fixed $r$ is much
greater than that of MIMO and MIMO antenna selection if the total
number of active ports or antennas is the same. 
\item Finally, we introduce a new performance metric, referred to as $q$-outage
capacity, that jointly considers both rate and outage probability.
We show that MIMO-FAS outperforms the traditional MIMO and MIMO antenna
selection in terms of $q$-outage capacity. This result suggests that
MIMO-FAS is more reliable in delivering high data rate transmission
than the traditional MIMO systems. 
\end{itemize}

\subsection{Organization and Notations}

The remainder of this paper is organized as follows. Section II introduces
the system model of MIMO-FAS. Section III details the proposed QR
MIMO-FAS scheme that maximizes its rate at high SNR. The optimal DMT
of MIMO-FAS is analyzed in Section IV. Section V presents the numerical
results to compare MIMO-FAS with the traditional MIMO systems and
we conclude the paper in Section VI.

\begin{table}[t]
\caption{The meanings of key notations.}
\centering{}%\begin{tabular}{c|p{6cm}|}
\begin{tabular}{r|p{5.5cm}}
\hline 
Notation  & \centering Meaning\tabularnewline
\hline 
\hline 
$\boldsymbol{A}$  & Activation port matrices at both of the transmitter and receiver sides\tabularnewline
\hline 
$\boldsymbol{A}_{s}$  & Activation port matrix at side $s$\tabularnewline
\hline 
$C_{{\rm sys}}^{q}$  & $q$-outage capacity of a system\tabularnewline
\hline 
$d\left(r\right)$  & Diversity gain for $r$ multiplexing gain\tabularnewline
\hline 
$\boldsymbol{G}$  & Circularly symmetric complex Gaussian matrix with i.i.d. entries\tabularnewline
\hline 
$\boldsymbol{H}$  & Complex channels of MIMO-FAS\tabularnewline
\hline 
$\overset{}{\bar{\boldsymbol{H}}}$  & Complex channels of the activated ports\tabularnewline
\hline 
$\boldsymbol{H}_{a}$  & Partial channels of the active ports\tabularnewline
\hline 
$\boldsymbol{H}_{i}$  & Partial channels of the inactive ports\tabularnewline
\hline 
$\boldsymbol{J}_{s}$  & Spatial correlation matrix at side $s$\tabularnewline
\hline 
$\hat{\boldsymbol{J}}_{s}$  & Approximated matrix of $\boldsymbol{J}_{s}$\tabularnewline
\hline 
$\boldsymbol{J}_{{\rm red}}^{s}$  & Full rank spatial correlation matrix at side $s$\tabularnewline
\hline 
$\boldsymbol{K}$  & Input covariance\tabularnewline
\hline 
$n_{s}$  & Total number of active ports\tabularnewline
\hline 
$n_{{\rm {max}}}/n_{{\rm {min}}}$  & Maximum/minimum of $n_{rx}$ and $n_{tx}$\tabularnewline
\hline 
$N_{{\rm {max}}}/N_{{\rm {min}}}$  & Maximum/minimum of $N_{rx}$ and $N_{tx}$\tabularnewline
\hline 
$N_{s}$  & Total number of ports at side $s$\tabularnewline
\hline 
$N_{i}^{s}$  & Number of ports in the $i$-th dimension at side $s$\tabularnewline
\hline 
$N_{s}^{'}$  & Rank of $\boldsymbol{J}_{{\rm red}}^{s}$\tabularnewline
\hline 
$N_{\min}^{'}$  & Minimum of $N_{rx}^{'}$ and $N_{tx}^{'}$\tabularnewline
\hline 
$P_{{\rm sys}}^{{\rm out}}\left({\rm SNR},r\right)$  & Outage probability of a system in terms of $r\log{\rm SNR}$\tabularnewline
\hline 
$\bar{P}_{{\rm sys}}^{{\rm out}}\left({\rm SNR},q\right)$  & Outage probability of a system for a fixed $q$-transmission rate\tabularnewline
\hline 
$\boldsymbol{P}$  & Power allocation matrix\tabularnewline
\hline 
$r$  & Multiplexing gain\tabularnewline
\hline 
$R_{{\rm sys}}\left({\rm SNR}\right)$  & Rate of a system for a given ${\rm SNR}$\tabularnewline
\hline 
$s\in\left\{ tx,rx\right\} $  & Subscript/superscript to denote the transmit/receiver side\tabularnewline
\hline 
${\rm SNR}$  & Transmit SNR\tabularnewline
\hline 
$\boldsymbol{v}_{l}^{s}$  & Certificate between the linear transformation of $\boldsymbol{J}_{s}^{\left(l\right)}$
and $\boldsymbol{J}_{{\rm sub}}^{\left(l-1\right)}$\tabularnewline
\hline 
$W_{s}$  & Total area of the fluid antenna at side $s$\tabularnewline
\hline 
$W_{i}^{s}$  & Length of the fluid antenna in the $i$-th dimension at side $s$\tabularnewline
\hline 
$\boldsymbol{W}$  & Transmit and receive beamforming matrices\tabularnewline
\hline 
$\boldsymbol{W}_{s}$  & Beamforming matrix at $s$ side\tabularnewline
\hline 
$\lambda$  & Wavelength of the carrier frequency\tabularnewline
\hline 
$\boldsymbol{\Lambda}_{s}$  & Matrix whose diagonal entries are the eigenvalues of $\boldsymbol{J}_{s}$\tabularnewline
\hline 
$\boldsymbol{\Sigma}$  & Matrix whose diagonal entries are singular values of $\boldsymbol{H}$\tabularnewline
\hline 
$\overset{}{\bar{\boldsymbol{\Sigma}}}$  & Matrix whose diagonal entries are singular values of $\boldsymbol{\bar{H}}$\tabularnewline
\hline 
$\boldsymbol{\Omega}$  & Special matrix used for port optimization\tabularnewline
\hline 
\end{tabular}\label{tab:notations} 
\end{table}

Throughout this paper, scalar variables are denoted by italic letters
(e.g., $c$), vectors are denoted by boldface italic small letters
(e.g., $\boldsymbol{c}$) and matrices are denoted by boldface italic
capital letters (e.g., $\boldsymbol{C}$). Additionally, $\left(\cdot\right)^{T}$
denotes transpose, $\left(\cdot\right)^{H}$ denotes conjugate transpose
while ${\rm det}\left(\cdot\right)$, ${\rm rank\left(\cdot\right)}$
and ${\rm trace}\left(\cdot\right)$ represent the determinant, rank
and trace of a matrix, respectively. Moreover, $\left|\cdot\right|$,
$\left\Vert \cdot\right\Vert _{2}$ and $\left\Vert \cdot\right\Vert _{F}$
denote the absolute, Euclidean norm and Frobenius norm operations,
respectively. Furthermore, we use $\log(\cdot)$ to denote logarithm
with base 2, $\left[\cdot\right]_{c}^{+}$ outputs the argument that
is lower bounded by $c$, $\min\left\{ \cdot\right\} $ and $\max\left\{ \cdot\right\} $
denote the minimum and maximum value of the argument, respectively.
${\mathbb{E}}[\cdot]$ returns the expected value of the input random
quantity, and $\otimes$ denotes the Kronecker product. Finally, $\boldsymbol{e}_{k}$
represents an all-zero vector except the $k$-th entry being unity,
${\rm diag}(\cdot)$ denotes a diagonal matrix whose diagonal entries
are the inputs, and $(\cdot)^{\dag}$ denotes the pseudoinverse of
an input matrix. To help readers follow the mathematical contents,
the meanings of the key variables are listed in Table \ref{tab:notations}.

\section{System Model}

\begin{figure*}
\centering{} \includegraphics[width=0.9\linewidth]{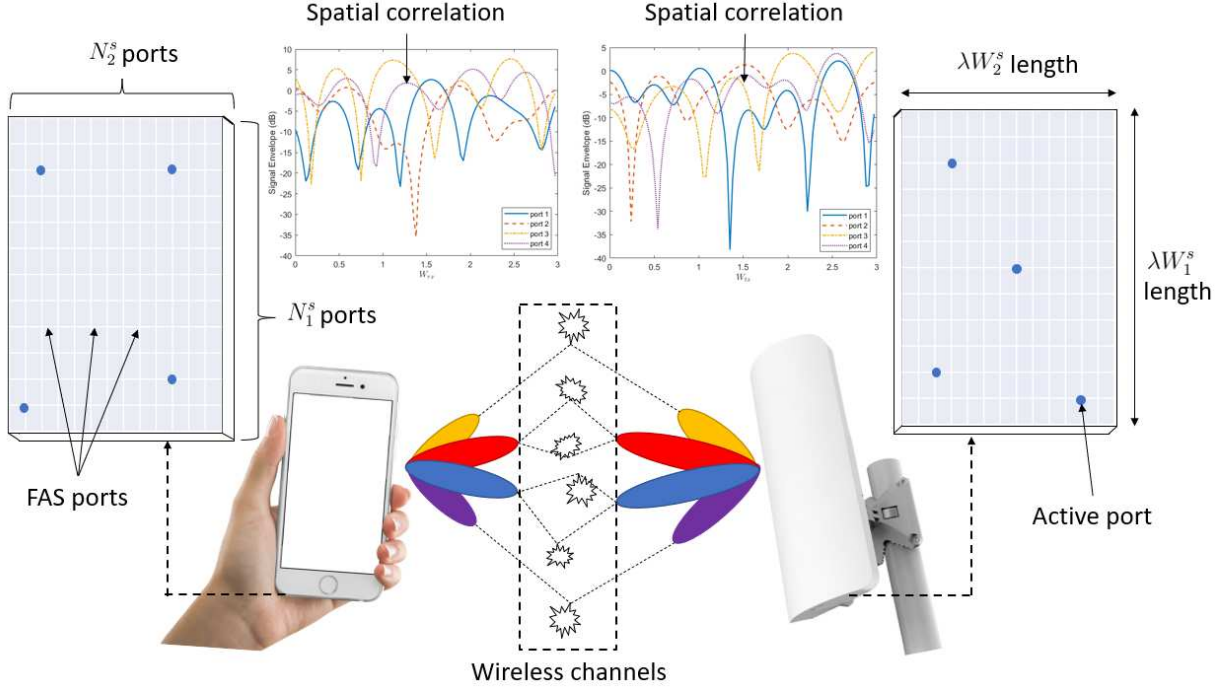}
\caption{A schematic of point-to-point MIMO-FAS.}
\label{fig:system} 
\end{figure*}

As illustrated in Fig.~\ref{fig:system}, we consider a point-to-point
wireless communication channel in which the transmitter and receiver
are equipped with a fluid antenna. To facilitate our discussions,
we use the subscript/superscript $s$ to denote the parameters at
the transmitter or receiver as $tx$ or $rx$, respectively, i.e.,
$s\in\left\{ tx,rx\right\} $. We assume that the fluid antenna takes
up a 2D space with an area of $W_{s}$ and has $N_{s}$ ports spread
uniformly over the 2D space. A grid structure is considered where
$N_{i}^{s}$ ports are uniformly distributed along a linear space
of length $\lambda W_{i}^{s}$ for $i\in\left\{ 1,2\right\} $, so
that $N_{s}=N_{1}^{s}\times N_{2}^{s}$ and $W_{s}=\lambda W_{1}^{s}\times\lambda W_{2}^{s}$,
where $\lambda$ is the wavelength of the carrier frequency. In this
MIMO-FAS, the transmitter and receiver can only activate $n_{s}$
out of $N_{s}$ ports. Note that for SISO-FAS, $n_{s}=1,\forall s$.

\textcolor{black}{Considering a 3D environment under rich scattering,
the spatial correlation between the $\left(n_{1}^{s},n_{2}^{s}\right)$-th
port and the $\left(\tilde{n}_{1}^{s},\tilde{n}_{2}^{s}\right)$-th
port is given by \vspace{-0.2cm}
 } \textcolor{black}{{} 
\begin{align}
 & J_{\left(n_{1}^{s},n_{2}^{s}\right),\left(\tilde{n}_{1}^{s},\tilde{n}_{2}^{s}\right)}^{s}\nonumber \\
 & =j_{0}\left(2\pi\sqrt{\left({\textstyle \frac{\left|n_{1}^{s}-\tilde{n}_{1}^{s}\right|}{N_{1}^{s}-1}W_{1}^{s}}\right)^{2}+\left({\textstyle \frac{\left|n_{2}^{s}-\tilde{n}_{2}^{s}\right|}{N_{2}^{s}-1}W_{2}^{s}}\right)^{2}}\right),\label{eq:1}
\end{align}
where $j_{0}\left(\cdot\right)$ is the spherical Bessel function
of the first kind.}\footnote{\textcolor{black}{Note that (\ref{eq:1}) can be reduced to a 1D fluid
antenna under 2D scattering environments by setting $N_{1}^{s}=1$
and $\frac{0}{0}\triangleq0$ and replacing $j_{0}\left(\cdot\right)$
by $J_{0}\left(\cdot\right)$ where $J_{0}\left(\cdot\right)$ is
the Bessel function of the first kind.}}\textcolor{black}{{} }A detailed proof can be found in Appendix I.
As seen in (\ref{eq:1}), it is cumbersome to label a port in 2D.
To simplify our notations, we use the function ${\rm map}:\mathbb{R}^{2}\rightarrow\mathbb{R}$,
e.g., ${\rm map}\left(n_{1}^{s},n_{2}^{s}\right)=l_{s}$, where $l_{s}\in\left\{ 1,\dots,N_{s}\right\} $.

Using the mapping function, we can express the spatial correlation
matrix $\boldsymbol{J}_{s}$ as 
\begin{equation}
\boldsymbol{J}_{s}=\left[\begin{array}{cccc}
J_{1,1}^{s} & J_{1,2}^{s} & \ldots & J_{1,N_{s}}^{s}\\
J_{2,1}^{s} & J_{2,2}^{s} & \ldots & J_{2,N_{s}}^{s}\\
\vdots & \vdots & \ddots & \vdots\\
J_{N_{s},1}^{s} & J_{N_{s},2}^{s} & \ldots & J_{N_{s},N_{s}}^{s}
\end{array}\right],\label{eq:2}
\end{equation}
where $J_{k_{s},l_{s}}^{s}$ is the spatial correlation of the $k_{s}$-th
and the $l_{s}$-th port at side $s$, and $k_{s}$ and $l_{s}$ are
the labels after mapping. Since $J_{k_{s},l_{s}}^{s}=J_{l_{s},k_{s}}^{s}$,
(\ref{eq:2}) can be decomposed into $\boldsymbol{J}_{s}=\boldsymbol{U}_{s}\boldsymbol{\Lambda}_{s}\boldsymbol{U}_{s}^{H}$
where $\boldsymbol{U}_{s}$ is an $N_{s}\times N_{s}$ matrix whose
$l$-th column (i.e., $\boldsymbol{u}_{l}^{s}$) is the eigenvector
of $\boldsymbol{J}_{s}$ and $\boldsymbol{\Lambda}_{s}={\rm diag}\left(\lambda_{1}^{s},\ldots,\lambda_{N_{s}}^{s}\right)$
is an $N_{s}\times N_{s}$ diagonal matrix whose $l$-th diagonal
entries are the corresponding eigenvalues of $\boldsymbol{u}_{l}^{s}$.
Without loss of generality, we assume that the values of the eigenvalues
in $\boldsymbol{\Lambda}_{s}$ are in descending order, i.e., $\lambda_{1}^{s}\geq\cdots\geq\lambda_{N_{s}}^{s}$.
Note that $\boldsymbol{U}_{s}\boldsymbol{\Lambda}_{s}\boldsymbol{U}_{s}^{H}$
is computed independently for each $s\in\left\{ tx,rx\right\} $.

\textcolor{black}{Given $\boldsymbol{U}_{s}$ and $\boldsymbol{\Lambda}_{s}$
for $\forall s$, the complex channel of MIMO-FAS can be modelled
as 
\begin{equation}
\boldsymbol{H}=\delta\:\boldsymbol{U}_{rx}\sqrt{\boldsymbol{\Lambda}_{rx}}\boldsymbol{G}\sqrt{\boldsymbol{\Lambda}_{tx}^{H}}\boldsymbol{U}_{tx}^{H},\label{eq:3}
\end{equation}
where $\boldsymbol{G}=\left[\boldsymbol{g}_{1},\ldots,\boldsymbol{g}_{N_{tx}}\right]$,
$\boldsymbol{g}_{l}=\left[g_{1,l},\ldots,g_{N_{rx},l}\right]^{T}$,
$g_{k,l}=x_{k,l}+jy_{k,l}$, and $x_{k,l},y_{k,l}$ are independent
Gaussian random variables with zero mean and variance of $\frac{1}{2}$,
$\forall k,l$, and $\delta^{2}$ is the path loss. Letting $\boldsymbol{h}_{{\rm vec}}={\rm vec}\left(\boldsymbol{H}\right)$,
the covariance of $\boldsymbol{h}_{{\rm vec}}$ is $\delta^{2}\left(\boldsymbol{J}_{tx}^{T}\otimes\boldsymbol{J}_{rx}\right)$,
where $\boldsymbol{J}_{tx}^{T}\otimes\boldsymbol{J}_{rx}$ is the
spatial correlation matrix between the transmit and receive ports.}\footnote{Due to the port spatial correlation, it can be shown that only a small
number of observed ports/training is required to obtain the full channel
state information regardless of the number of ports of the FAS \cite{WongIL}.
Machine learning techniques have also been proposed to address the
channel estimation problem for port selection in FAS, e.g., \cite{10018377}.}

Next, we denote the activation port matrices at the transmitter and
receiver, respectively, as $\boldsymbol{A}_{tx}=\left[\boldsymbol{a}_{1}^{tx},\ldots,\boldsymbol{a}_{n_{tx}}^{tx}\right]$
and $\boldsymbol{A}_{rx}=\left[\boldsymbol{a}_{1}^{rx},\ldots,\boldsymbol{a}_{n_{rx}}^{rx}\right]^{T}$,
where $\boldsymbol{a}_{l}^{tx}$ and $\boldsymbol{a}_{l}^{rx}$ are
standard basis vector (i.e., $\boldsymbol{a}_{l}^{s}\in\left\{ \boldsymbol{e}_{1},\ldots,\boldsymbol{e}_{N_{s}}\right\} $).
Since only distinct $n_{s}$ ports can be activated at a time, we
have $\boldsymbol{a}_{k}^{tx}\neq\boldsymbol{a}_{l}^{tx}$ and $\boldsymbol{a}_{k}^{rx}\neq\boldsymbol{a}_{l}^{rx}$
if $k\neq l$. Let us further denote $\boldsymbol{W}_{tx}$ and $\boldsymbol{W}_{rx}$
as the transmit and receive beamforming matrices, with the constraint
$\left\Vert \boldsymbol{W}_{s}\right\Vert _{2}=1$. Then the receive
signals of MIMO-FAS can be rewritten as 
\begin{align}
\boldsymbol{W}_{rx}\boldsymbol{A}_{rx}\boldsymbol{Y} & =\boldsymbol{W}_{rx}\boldsymbol{A}_{rx}\boldsymbol{H}\boldsymbol{A}_{tx}\boldsymbol{W}_{tx}\boldsymbol{x}+\boldsymbol{W}_{rx}\boldsymbol{A}_{rx}\boldsymbol{w}\label{eq:4}\\
\Rightarrow\tilde{\boldsymbol{Y}} & =\tilde{\boldsymbol{H}}\boldsymbol{x}+\tilde{\boldsymbol{w}},\label{eq:5}
\end{align}
where $\boldsymbol{x}$ is the information signal and $\boldsymbol{w}$
is the additive white Gaussian noise with zero mean and identity covariance.
In (\ref{eq:5}), we defined $\tilde{\boldsymbol{Y}}\triangleq\boldsymbol{W}_{rx}\boldsymbol{A}_{rx}\boldsymbol{Y}$,
$\tilde{\boldsymbol{H}}\triangleq\boldsymbol{W}_{rx}\boldsymbol{A}_{rx}\boldsymbol{H}\boldsymbol{A}_{tx}\boldsymbol{W}_{tx}$
and $\tilde{\boldsymbol{w}}\triangleq\boldsymbol{W}_{rx}\boldsymbol{A}_{rx}\boldsymbol{w}$.
For ease of expositions, we denote $\boldsymbol{A}=\left[\boldsymbol{A}_{tx},\boldsymbol{A}_{rx}\right]$,
$\boldsymbol{W}=\left[\boldsymbol{W}_{tx},\boldsymbol{W}_{rx}\right]$,
$\bar{\boldsymbol{H}}=\boldsymbol{A}_{rx}\boldsymbol{H}\boldsymbol{A}_{tx}$
and $\boldsymbol{K}=\boldsymbol{W}_{tx}\boldsymbol{P}\boldsymbol{W}_{tx}^{H}$
where $\boldsymbol{K}$ is the input covariance and $\boldsymbol{P}=\mathbb{E}\left[\boldsymbol{x}\boldsymbol{x}^{H}\right]$
is the power allocation matrix. Then, the rate of MIMO-FAS is given
by 
\begin{equation}
R\left(\boldsymbol{A},\boldsymbol{W},\boldsymbol{P}\right)=\log{\rm det}\left(\boldsymbol{I}+\bar{\boldsymbol{H}}\boldsymbol{K}\bar{\boldsymbol{H}}^{H}\right),\label{eq:6}
\end{equation}
where ${\rm trace}\left(\boldsymbol{K}\right)\leq{\rm SNR}$ and ${\rm SNR}$
is the transmit SNR.

\section{QR MIMO-FAS: Suboptimal Port Selection, Beamforming and Power Allocation}

In this section, we aim to maximize the rate of the MIMO-FAS via joint
optimal port selection, beamforming and power allocation. The optimization
problem is formulated as 
\begin{subequations}
\label{OP_1} 
\begin{align}
\max_{\boldsymbol{A},\boldsymbol{W},\boldsymbol{P}}~~ & R\left(\boldsymbol{A},\boldsymbol{W},\boldsymbol{P}\right)\label{eq:7}\\
{\rm s.t.}~~ & \boldsymbol{a}_{l}^{s}\in\left\{ e_{1},\ldots,e_{N_{s}}\right\} ,s\in\left\{ rx,tx\right\} ,\forall l,\label{eq:8}\\
 & \boldsymbol{a}_{k}^{tx}\neq\boldsymbol{a}_{l}^{tx},\text{\,}\text{{if}}\,k\neq l,\label{eq:9}\\
 & \boldsymbol{a}_{k}^{rx}\neq\boldsymbol{a}_{l}^{rx},\,\text{{if}}\,k\neq l,\label{eq:10}\\
 & \left\Vert \boldsymbol{W}_{tx}\right\Vert _{2}=\left\Vert \boldsymbol{W}_{rx}\right\Vert _{2}=1,\label{eq:11}\\
 & {\rm trace}\left(\boldsymbol{K}\right)\leq{\rm SNR}.\label{eq:12}
\end{align}
\end{subequations}
 Note that \eqref{OP_1} is a non-convex optimization problem because
i) the optimization variables are mutually coupled and ii) its domain
is non-convex. A systematic way to solve this problem is to employ
an exhaustive search \cite{1341263}, SVD and waterfilling power allocation
\cite{1203154}. In particular, the maximum rate of MIMO-FAS can be
computed via SVD and waterfilling power allocation over ${\scriptstyle \left(\begin{array}{c}
N_{tx}\\
n_{tx}
\end{array}\right)}\times{\scriptstyle \left(\begin{array}{c}
N_{rx}\\
n_{rx}
\end{array}\right)}$ port combinations.\footnote{As will be shown later in this paper, $N_{s},\forall s$ can be represented
by a finite constant even in cases where $N_{s}\rightarrow\infty$.} Nevertheless, such a method requires a non-polynomial time complexity
of $\mathcal{O}\left(N_{tx}^{n_{tx}}N_{rx}^{n_{rx}}\right)$ which
is prohibitively high.

To reduce the time complexity, we propose a suboptimal scheme, namely
QR MIMO-FAS, to maximize the rate of MIMO-FAS in the high SNR regime.
This scheme is useful for analyzing the DMT of MIMO-FAS. In the proposed
scheme, we decouple \eqref{OP_1} into two subproblems: i) optimal
port selection and ii) optimal beamforming and power allocation. For
optimal port selection, \eqref{OP_1} can be simplified as 
\begin{equation}
\max_{\boldsymbol{A}}R\left(\boldsymbol{A},\boldsymbol{W},\boldsymbol{P}\right)~~{\rm s.t.}~~(\ref{eq:8}),(\ref{eq:9}),(\ref{eq:10}).\label{eq:13}
\end{equation}
However, it is still challenging to solve (\ref{eq:13}) since the
objective function cannot be directly evaluated. To overcome this
problem, we exploit the fact that $R\left(\boldsymbol{A},\boldsymbol{W},\boldsymbol{P}\right)$
strongly depends on ${\rm det}(\bar{\boldsymbol{H}}\bar{\boldsymbol{H}}^{H})$
in the high SNR regime \cite{7343424}. Thus, (\ref{eq:13}) can be
relaxed as 
\begin{equation}
\max_{\boldsymbol{A}}{\rm det}(\bar{\boldsymbol{H}}\bar{\boldsymbol{H}}^{H})~~{\rm s.t.}~~(\ref{eq:8}),(\ref{eq:9}),(\ref{eq:10}),\label{eq:14}
\end{equation}
which is unfortunately still an NP-hard problem \cite{CIVRIL20094801}.\footnote{The relaxation is done because $R\left(\boldsymbol{A},\boldsymbol{W},\boldsymbol{P}\right)\approx\log\det(\bar{\boldsymbol{H}}\boldsymbol{K}\bar{\boldsymbol{H}}^{H})$
at high SNR and the rate of using equal power allocation approaches
to that of waterfilling power allocation as SNR increases \cite{5733445}.
For other SNR regimes, solving \eqref{OP_1} with low complexity remains
open. Nevertheless, we can obtain an efficient solution at low SNR
by activating $n_{rx}/n_{tx}$ ports where the row/column-norm of
$\boldsymbol{H}$ are the largest and they are separated by at least
$c_{rx}/c_{tx}$ distance. We refer this scheme as the greedy selection.} However, we can obtain a suboptimal solution by using the strong
rank-revealing QR (RRQR) factorization \cite{doi:10.1137/0917055}.

Specifically, by applying QR factorization with pivoted column on
$\boldsymbol{H}^{H}$, we have 
\begin{equation}
\boldsymbol{H}^{H}\boldsymbol{\varPi}=\boldsymbol{Q}\boldsymbol{R},\label{eq:15}
\end{equation}
where $\boldsymbol{\varPi}$ is a permutation matrix, $\boldsymbol{Q}$
is an orthogonal matrix and $\boldsymbol{R}$ is an upper triangular
matrix where the absolute of leading entries in $\boldsymbol{R}$
are decreasing in values. The upper triangular matrix $\boldsymbol{R}$
can be rewritten as 
\begin{equation}
\boldsymbol{R}=\left[\boldsymbol{R}_{1}~\boldsymbol{R}_{2}\right]=\left[\begin{array}{cc}
\boldsymbol{S}_{1} & \boldsymbol{S}_{2}\\
\boldsymbol{0} & \boldsymbol{S}_{3}
\end{array}\right],\label{eq:16}
\end{equation}
where $\boldsymbol{R_{1}}=\left[\boldsymbol{S}_{1}~\boldsymbol{0}\right]^{T}$
and $\boldsymbol{R}_{2}=\left[\boldsymbol{S}_{2}~\boldsymbol{S}_{3}\right]^{T}$.
Substituting (\ref{eq:16}) into (\ref{eq:15}), we obtain 
\begin{equation}
\left[\boldsymbol{H}^{H}\boldsymbol{\varPi}_{a}~\boldsymbol{H}^{H}\boldsymbol{\varPi}_{i}\right]=\left[\boldsymbol{H}_{a}^{H}~\boldsymbol{H}_{i}^{H}\right]=\left[\boldsymbol{Q}\boldsymbol{R}_{1}~\boldsymbol{Q}\boldsymbol{R}_{2}\right],\label{eq:17}
\end{equation}
in which $\boldsymbol{\varPi}=\left[\boldsymbol{\varPi}_{a}~\boldsymbol{\varPi}_{i}\right]$.
In (\ref{eq:17}), the left hand side and right hand side of (\ref{eq:15})
are separated into two blocks and thus we can interpret $\boldsymbol{H}_{a}^{H}$
and $\boldsymbol{H}_{i}^{H}$ as the MIMO-FAS channels of active and
inactive ports, respectively. Since the singular values of $\boldsymbol{H}^{H}$
and $\boldsymbol{R}$ remain the same, it is clear that (\ref{eq:17})
provides the following properties 
\begin{align}
\prod_{m=1}^{N_{\min}}\sigma_{m}\left(\boldsymbol{H}^{H}\right) & =\prod_{m=1}^{n_{s}}\sigma_{m}\left(\boldsymbol{H}_{a}^{H}\right)\prod_{m=n_{s}+1}^{N_{\min}}\sigma_{m}\left(\boldsymbol{H}_{i}^{H}\right)\nonumber \\
 & =\prod_{m=1}^{n_{s}}\sigma_{m}\left(\boldsymbol{R}_{1}\right)\prod_{m=n_{s}+1}^{N_{\min}}\sigma_{m}\left(\boldsymbol{R}_{2}\right),\label{eq:18}
\end{align}
where $N_{\min}=\min\left\{ N_{rx},N_{tx}\right\} $ and $\sigma_{m}\left(\cdot\right)$
denotes the $m$-th singular value of the matrix argument.

In alignment with (\ref{eq:14}), our objective here is to maximize
$\prod_{m=1}^{n_{s}}\sigma_{m}\left(\boldsymbol{H}_{a}^{H}\right)$
in (\ref{eq:18}) by permuting the $k$-th column of $\boldsymbol{H}_{a}^{H}$
and the $l$-th column of $\boldsymbol{H}_{i}^{H}$. To facilitate
this objective, we employ the matrix $\boldsymbol{\Omega}$ where
the $(k,l)$-th entry of $\boldsymbol{\Omega}$ is 
\begin{equation}
\Omega_{k,l}=\sqrt{\left|\boldsymbol{S}{}_{1}^{\dagger}\boldsymbol{S}_{2}\right|_{k,l}^{2}+\left\Vert \boldsymbol{s}_{3,l}\right\Vert _{2}^{2}+\left\Vert \boldsymbol{s}_{1,k}^{\dagger,T}\right\Vert _{2}^{2}},\label{eq:19}
\end{equation}
where $\left|\boldsymbol{S}\right|_{k,l}$ gives the absolute value
of the $(k,l)$-th entry of $\boldsymbol{S}$, $\boldsymbol{s}_{s,l}$
is the $l$-th column of $\boldsymbol{S}_{s}$ and $\boldsymbol{s}_{s,k}^{\dagger,T}$
is the $k$-th row of $\boldsymbol{S}_{s}^{\dagger}$. Furthermore,
let us denote $\boldsymbol{H}_{a,k,l}^{H}$ (or $\boldsymbol{\varPi}_{a,k,l}$)
is the new matrix where the $k$-th column of $\boldsymbol{H}_{a}^{H}$
(or $\boldsymbol{\varPi}_{a}$) and the $l$-th column of $\boldsymbol{H}_{i}^{H}$
(or $\boldsymbol{\varPi}_{i}$) are permuted.

Conventionally, it is necessary to permute all the $\left(k,l\right)$
combinations and find the maximum $\prod_{m=1}^{n_{s}}\sigma_{m}\left(\boldsymbol{H}_{a,k,l}^{H}\right)$
in the presence of spatial correlation. Nevertheless, using (\ref{eq:19}),
we can determine the increase or decrease of $\prod_{m=1}^{n_{s}}\sigma_{m}\left(\boldsymbol{H}_{a,k,l}^{H}\right)$
over $\prod_{m=1}^{n_{s}}\sigma_{m}\left(\boldsymbol{H}_{a}^{H}\right)$
before permuting them. Hence, we can directly permute the $k$-th
column of $\boldsymbol{H}_{a}^{H}$ (and $\boldsymbol{\varPi}_{a}$)
and the $l$-th column of $\boldsymbol{H}_{i}^{H}$ (and $\boldsymbol{\varPi}_{i}$)
and then update $\boldsymbol{H}^{H}=\boldsymbol{H}_{a,k,l}^{H}$ (and
$\boldsymbol{\varPi}=\boldsymbol{\varPi}{}_{a,k,l}$) if $\Omega_{k,l}>1$.
In this paper, we permute the columns based on the largest $\Omega_{k,l}$
which yields a suboptimal solution. Furthermore, we perform QR factorization
to obtain the expression in (\ref{eq:15}) with the existing $\boldsymbol{\varPi}$.
These steps can be repeated until $\Omega_{k,l}\leq1~\forall k,l$.
Using strong RRQR factorization, we can obtain $\boldsymbol{H}_{a}^{H}$
with the maximum $\prod_{m=1}^{n_{s}}\sigma_{m}\left(\boldsymbol{H}_{a}^{H}\right)$
where $\boldsymbol{A}_{rx}^{*}=\boldsymbol{\varPi}_{a}^{T}$. Reapplying
strong RRQR factorization on $\boldsymbol{A}_{rx}^{*}\boldsymbol{H}$,
we can also obtain $\boldsymbol{A}_{tx}^{*}$ where $\boldsymbol{A}_{tx}^{*}=\boldsymbol{\varPi}_{a}$.

Given $\boldsymbol{A}^{*}$, \eqref{OP_1} reduces to the optimal
beamforming and power allocation problem which can be formulated as
\begin{equation}
\max_{\boldsymbol{W},\boldsymbol{P}}R\left(\boldsymbol{W},\boldsymbol{P}|\boldsymbol{A}^{*}\right)~~{\rm s.t.}~~(\ref{eq:11}),(\ref{eq:12}).\label{eq:20}
\end{equation}
Interestingly, (\ref{eq:20}) can be easily solved via SVD and waterfilling
power allocation \cite{1203154}. In particular, the solution to (\ref{eq:20})
is $\boldsymbol{W}_{r}^{*}=\bar{\boldsymbol{M}}^{H}$ and $\boldsymbol{W}_{t}^{*}=\bar{\boldsymbol{N}}$
where $\bar{\boldsymbol{H}}=\bar{\boldsymbol{M}}\bar{\boldsymbol{\Sigma}}\bar{\boldsymbol{N}}^{H}$,
$\bar{\boldsymbol{M}}$ is the left singular matrix of $\bar{\boldsymbol{H}}$,
$\bar{\boldsymbol{N}}$ is the right singular matrix of $\bar{\boldsymbol{H}}$,
$\bar{\boldsymbol{\Sigma}}={\rm diag}\left(\bar{\Sigma}_{1},\dots,\bar{\Sigma}_{n_{\min}}\right)$
denotes the diagonal matrix whose $l$-th entry is the $l$-th singular
value of $\bar{\boldsymbol{H}}$, $\bar{\Sigma}_{1}\geq\cdots\geq\bar{\Sigma}_{n_{\min}}$
and $n_{\min}=\min\left\{ n_{tx},n_{rx}\right\} $. In addition, $\boldsymbol{P}^{*}={\rm diag}\left(p_{1}^{*},\dots,p_{n_{\min}}^{*}\right)$,
$p_{l}^{*}=\left[\mu-\frac{1}{\overset{}{\bar{\Sigma}_{l}^{2}}}\right]_{0}^{+}$,
and ${\rm SNR}=\sum p_{l}^{*}$.\footnote{\textcolor{black}{Note that it is possible to employ equal power allocation
at high SNR. In fact, our analysis leverages this assumption for tractability.
Nevertheless, we will consider waterfilling power allocation here
since it is optimal regardless of the SNR. In addition, it would be
useful later to make a fair comparison between different benchmarking
schemes.}} Thus, the optimal input covariance is $\boldsymbol{K}^{*}=\boldsymbol{W}_{tx}^{*}\boldsymbol{P}^{*}\boldsymbol{W}_{tx}^{*H}$.
Using the above methods, the rate of the QR MIMO-FAS for a given ${\rm SNR}$
can be expressed as 
\begin{equation}
R_{{\rm QR}}\left({\rm SNR}\right)=\sum_{l=1}^{n_{\min}}\log\left(1+p_{l}^{*}\bar{\Sigma}_{l}^{2}\right),\label{eq:21}
\end{equation}
where (\ref{eq:21}) helps analyze the optimal DMT of MIMO-FAS.

\begin{algorithm}[t]
\begin{algorithmic}[1] \STATE Compute (\ref{eq:15}) via QR factorization
with pivoted column

\STATE Compute $\boldsymbol{\Omega}$ using (\ref{eq:19})

\STATE \textbf{While} $\Omega_{k,l}>1$

\STATE Permute the $k$-th column of $\boldsymbol{H}_{a}^{H}/\boldsymbol{\varPi}_{a}$
and the $l$-th column of $\boldsymbol{H}_{i}^{H}/\boldsymbol{\varPi}_{i}$
with the largest $\Omega_{k,l}>1$

\STATE Update $\boldsymbol{H}^{H}=\boldsymbol{H}_{a,k,l}^{H}$ and
$\boldsymbol{\varPi}=\boldsymbol{\varPi}{}_{a,k,l}$

\STATE Perform QR factorization using $\boldsymbol{\varPi}$ and
return to Step 2

\STATE \textbf{end}

\STATE Set $\boldsymbol{A}_{r}^{*}=\boldsymbol{\varPi}_{a}^{T}$

\STATE Repeat Steps 1--7 by replacing $\boldsymbol{H}^{H}$ with
$\boldsymbol{A}_{r}^{*}\boldsymbol{H}$

\STATE Set $\boldsymbol{A}_{r}^{*}=\boldsymbol{\varPi}_{a}^{T}$

\STATE Compute SVD on $\boldsymbol{A}_{r}^{*}\boldsymbol{H}\boldsymbol{A}_{t}^{*}$
to obtain $\boldsymbol{W}^{*}$

\STATE Use bisection to obtain $\boldsymbol{P}^{*}$ \end{algorithmic}
\caption{Pseudocode of QR MIMO-FAS}
\label{alg:1} 
\end{algorithm}

The pseudocode of the proposed scheme is presented in Algorithm \ref{alg:1}.
Let us denote $N_{\max}=\max\left\{ N_{tx},N_{rx}\right\} $ and $n_{\max}=\max\left\{ n_{tx},n_{rx}\right\} $.
The worst computational cost of Step 1 is $\frac{4}{3}n_{\max}^{3}-4N_{\max}n_{\max}^{2}+4N_{\max}^{2}n_{\max}$
flops \cite{doi:10.1137/0917055}. Steps 2--7 require $\left(1+t_{i}\right)\left[\left(\frac{2}{3}n_{\max}^{3}N_{\max}^{2}+2n_{\max}^{2}N_{\max}^{3}+2N_{\max}^{3}\right)\right]$
flops \cite{ford2014numerical}, where $t_{i}$ is a finite number
of permutations and it is usually small due to Step 1 \cite{7343424}.
Step 9 has the same total computational cost as Steps 1--7, which
is also finite. In addition, Steps 11 and 12 require $21n_{\max}^{3}$
and $\log\left(\frac{\mu_{\max}}{\epsilon_{0}}\right)n_{\max}$ flops,
respectively, where $\mu_{\max}$ is the interval for searching $\mu$
and $\epsilon_{0}$ is the tolerance for bisection method. Summing
up the computational costs, the proposed scheme has a polynomial time
complexity of $\mathcal{{O}}\left(n_{\max}^{3}N_{\max}^{3}\right)$.
Compared to the global optimal solution, the proposed scheme significantly
reduces the time complexity.

\section{Optimal DMT}

In this section, we analyze the optimal DMT of MIMO-FAS. As defined
in \cite{1197843}, a MIMO scheme is said to achieve a multiplexing
gain of $r$ and a diversity gain of $d$ if 
\begin{equation}
\lim_{{\rm SNR}\rightarrow\infty}\frac{R_{{\rm sys}}\left({\rm SNR}\right)}{\log{\rm SNR}}=r,\label{eq:22}
\end{equation}
and the outage probability satisfies\footnote{Here, we use the fact that the error probability can be arbitrarily
close to the outage probability \cite{4016314,1221802}.} 
\begin{equation}
\lim_{{\rm SNR}\rightarrow\infty}\frac{\log\left(P_{{\rm sys}}^{{\rm out}}\left({\rm SNR},r\right)\right)}{\log{\rm SNR}}=-d\left(r\right),\label{eq:23}
\end{equation}
in which $R_{{\rm sys}}\left({\rm SNR}\right)$ and $P_{{\rm sys}}^{{\rm out}}\left({\rm SNR},r\right)$
are, respectively, the rate and outage probability of the system.
Similar to \cite{1197843}, we use the symbol $\dot{=}$ to denote
exponential equality. In particular, $f\left({\rm SNR}\right)\dot{=}~{\rm SNR}^{-d}$
if 
\begin{equation}
\lim_{{\rm SNR}\rightarrow\infty}\frac{\log\left(f\left({\rm SNR}\right)\right)}{\log{\rm SNR}}=-d.\label{eq:24}
\end{equation}
To obtain the optimal DMT of MIMO-FAS, we present the following lemmas.
\begin{lem}
\label{lem:lemma1} \textcolor{black}{{}} Given a 2D space with an
area of $W_{s}$ where both $W_{1}^{s}\gg0$ and $W_{2}^{s}\gg0$,
$\boldsymbol{J}_{s}$ in (\ref{eq:2}) can be represented by $\boldsymbol{J}_{{\rm red}}^{s}$,
i.e., a full rank symmetric $N_{s}^{'}\times N_{s}^{'}$ finite-size
matrix even if $N_{s}\rightarrow\infty$. 
\end{lem}
\begin{IEEEproof}
For a positive $W_{s}$, consider (\ref{eq:2}) where $N_{s}\rightarrow\infty$.
Without loss of generality, we focus on two neighboring ports: the
$\left(n_{1}^{s},n_{2}^{s}\right)$-th and $\left(\tilde{n}_{1}^{s},\tilde{n}_{2}^{s}\right)$-th
port. In cases where $N_{1}^{s}\rightarrow\infty$, we analyze $\left(\tilde{n}_{1}^{s},\tilde{n}_{2}^{s}\right)=\left(n_{1}^{s}\pm1,n_{2}^{s}\right)$.
The spatial correlation between the $\left(n_{1}^{s},n_{2}^{s}\right)$-th
and $\left(\tilde{n}_{1}^{s},n_{2}^{s}\right)$-th port is given by\textcolor{black}{{}
\begin{equation}
J_{\left(n_{1}^{s},n_{2}^{s}\right),\left(\tilde{n}_{1}^{s},n_{2}^{s}\right)}^{s}=\underset{N_{1}^{s}\rightarrow\infty}{\lim}j_{0}\left(\frac{2\pi}{N_{1}^{s}-1}W_{1}^{s}\right)=1,\label{eq:26}
\end{equation}
}since $\underset{N_{1}^{s}\rightarrow\infty}{\lim}\frac{1}{N_{1}^{s}-1}=0$.\textcolor{black}{{}
In other words, the spatial correlation of the $\tilde{n}_{1}^{s}$-th
port is identical to that of the $n_{1}^{s}$-th port in the limit.}
For ease of exposition, let us refer to the spatial correlation of
a port as an entry.

For $n_{2}^{s}=\left\{ 1,\ldots,N_{2}^{s}\right\} $, we can remove
the identical entries (e.g., $\tilde{n}_{1}^{s}$ for $\exists n_{1}^{s}$)
and obtain $\bar{N}_{1}^{s}$ distinct entries. \textcolor{black}{Next,
let us denote the $\left(\dot{n}_{1}^{s},n_{2}^{s}\right)$-th port
as the farthest port away from the $\left(n_{1}^{s},n_{2}^{s}\right)$-th
port. Since $W_{1}^{s}\gg0$, it is clear that $J_{\left(n_{1}^{s},n_{2}^{s}\right),\left(\dot{n}_{1}^{s},n_{2}^{s}\right)}^{s}=j_{0}\left(2\pi\dot{c}\right)$
where $\dot{c}\gg0$ is the distance between the $\left(\dot{n}_{1}^{s},n_{2}^{s}\right)$-th
and $\left(n_{1}^{s},n_{2}^{s}\right)$-th ports. By intermediate
value theorem, we can conclude there is an $\varepsilon$ such that
\begin{equation}
\varepsilon={\rm inf}\left\{ e\left|J_{\left(n_{1}^{s},n_{2}^{s}\right),\left(n_{1}^{s}\pm e,n_{2}^{s}\right)}^{s}\right.\neq1,\begin{array}{l}
e\in\mathbb{N},\\
0<e\leq N_{1}^{s}-1
\end{array}\right\} .\label{eq:21e}
\end{equation}
Note that one cannot make any assumption on $e$ and $\varepsilon$
except their existence. Let us define $c_{1}^{s}\triangleq\frac{\varepsilon}{N_{1}^{s}-1}W_{1}^{s}$
as the minimal distance required for the spatial correlation between
the $\left(n_{1}^{s},n_{2}^{s}\right)$-th and $\left(n_{1}^{s}\pm e,n_{2}^{s}\right)$-th
ports to be completely distinct. }Then we can verify that $\bar{N}_{1}^{s}$
is finite since $W_{1}^{s}\geq\bar{N}_{1}^{s}c_{1}^{s}>0$. Let us
write the $\bar{N}_{1}^{s}$ distinct entries as a vector $\boldsymbol{v}_{n_{2}^{s}}$
for each $n_{2}^{s}\in\left\{ 1,\ldots,N_{2}^{s}\right\} $. If these
vectors are linearly dependent, then we can similarly remove the dependent
vectors and obtain $\bar{N}_{2}$ independent vectors. As a result,
we can rewrite $\boldsymbol{J}_{s}$ as a symmetric $\bar{N}_{s}\times\bar{N}_{s}$
finite-size matrix with distinct entries where $\bar{N}_{s}=\bar{N}_{1}^{s}\bar{N}_{2}^{s}$.

Using a similar argument, we see that $\boldsymbol{J}_{s}$ can be
represented by an $\bar{N}_{s}\times\bar{N}_{s}$ finite-size matrix
if $N_{2}^{s}\rightarrow\infty$. Combining the two cases, it is straightforward
to see that $\boldsymbol{J}_{s}$ can be rewritten as a symmetric
$\bar{N}_{s}\times\bar{N}_{s}$ matrix as $N_{1}^{s}\rightarrow\infty$
and $N_{2}^{s}\rightarrow\infty$ since both $W_{1}^{s}\gg0$ and
$W_{2}^{s}\gg0$, and we have (\ref{eq:21e}) and 
\begin{equation}
\varepsilon_{2}={\rm inf}\left\{ e\left|J_{\left(n_{1}^{s},n_{2}^{s}\right),\left(n_{1}^{s},n_{2}^{s}\pm e\right)}^{s}\right.\neq1,\begin{array}{l}
e\in\mathbb{N},\\
0<e\leq N_{2}^{s}-1
\end{array}\right\} .\label{eq:22e}
\end{equation}
From the above, it is clear that we can use the same argument to show
that $\boldsymbol{J}_{s}$ can be rewritten as a symmetric $\bar{N}_{s}\times\bar{N}_{s}$
matrix if $N_{1}^{s}/N_{2}^{s}$ is finite since we can always remove
the entries where their vertical/horizontal distances between the
adjacent ports are less than $c_{1}^{s}/c_{2}^{s}$.

Let us denote $N_{s}^{'}$ as the full rank of the symmetric $\bar{N}_{s}\times\bar{N}_{s}$
matrix where $N_{s}^{'}\leq\bar{N}_{s}$. Then we can further reduce
the symmetric $\bar{N}_{s}\times\bar{N}_{s}$ matrix to a full rank
symmetric $N_{s}^{'}\times N_{s}^{'}$ submatrix $\boldsymbol{J}_{{\rm red}}^{s}$
by removing the $(\bar{N}_{s}-N_{s}^{'})$ dependent rows and columns.
Thus, $\boldsymbol{J}_{s}$ can always be represented by $\boldsymbol{J}_{{\rm red}}^{s}$,
i.e., a full rank symmetric $N_{s}^{'}\times N_{s}^{'}$ finite-size
matrix. In other words, it suffices to consider $\boldsymbol{J}_{{\rm red}}^{s}$
instead of $\boldsymbol{J}_{s}$ since some rows/columns of $\boldsymbol{J_{s}}$
are identical or a linear combination of the others. A more general
result is given in Appendix II. 
\end{IEEEproof}
\begin{lem}
\label{lem:lemma2} If $\boldsymbol{J}_{tx}$ and $\boldsymbol{J}_{rx}$
are full rank, the optimal DMT of any MIMO system with channel $\boldsymbol{H}$
is the same as that of a system with channel $\boldsymbol{G}$. 
\end{lem}
\begin{IEEEproof}
See \cite{4137906}. 
\end{IEEEproof}
\begin{lem}
\label{lem:lemma3} The optimal DMT of using only $n_{rx}\times n_{tx}$
channels from the MIMO channel $\boldsymbol{G}$, where $n_{tx}\leq N_{tx}$
and $n_{rx}\leq N_{rx}$, is a piecewise linear function connecting
the points $\left(n_{\min},0\right)$ and 
\begin{equation}
\left\{ r,\left(N_{rx}-r\right)\left(N_{tx}-r\right)\right\} ,\quad r=0,\ldots,N,\label{eq:27}
\end{equation}
where 
\begin{equation}
N=\arg\min_{{\eta\in\mathbb{Z}\atop 0\leq\eta\leq n_{\min}-1}}\frac{\left(N_{rx}-\eta\right)\left(N_{tx}-\eta\right)}{n_{\min}-\eta}.\label{eq:28}
\end{equation}
\end{lem}
\begin{IEEEproof}
See \cite{5288958}. 
\end{IEEEproof}
\begin{cor}
If the antennas are placed based on a grid structure with at least
half a wavelength apart and the transmit/receive spatial correlation
matrices are full rank, then the optimal DMT of $n_{rx}\times n_{tx}$
MIMO antenna selection is a piecewise linear function connecting the
points $\left(n_{\min},0\right)$ and 
\begin{equation}
\left\{ r,\left(w_{rx}-r\right)\left(w_{tx}-r\right)\right\} ,\quad r=0,\ldots,N_{\text{as}},\label{eq:28_A}
\end{equation}
where 
\begin{equation}
N_{\text{as}}=\arg\min_{{\eta\in\mathbb{Z}\atop 0\leq\eta\leq n_{\min}-1}}\frac{\left(w_{rx}-\eta\right)\left(w_{tx}-\eta\right)}{n_{\min}-\eta}.\label{eq:28_B}
\end{equation}
\end{cor}
\begin{IEEEproof}
Given a fixed $W_{s}^{1}$ and $W_{s}^{2}$, there can be at most
$w_{s}=\left(\left\lfloor \frac{W_{s}^{1}}{0.5}\right\rfloor +1\right)$$\left(\left\lfloor \frac{W_{s}^{2}}{0.5}\right\rfloor +1\right)$
antennas at side $s$. Using Lemma \ref{lem:lemma2} and Lemma \ref{lem:lemma3},
we obtain (\ref{eq:28_A}) and (\ref{eq:28_B}). 
\end{IEEEproof}
Using the above lemmas, we can now obtain the outer bound of the DMT
of MIMO-FAS.
\begin{thm}
\label{thm:thm4} For finite $W_{rx}$ and $W_{tx}$, the outer bound
of the DMT of MIMO-FAS is a piecewise linear function connecting the
points $\left(n_{\min},0\right)$ and 
\begin{equation}
\left\{ r,\left(N_{rx}^{'}-r\right)\left(N_{tx}^{'}-r\right)\right\} ,\quad r=0,\dots,N',\label{eq:30}
\end{equation}
where 
\begin{equation}
N'=\arg\min_{{\eta\in\mathbb{Z}\atop 0\leq\eta\leq n_{\min}-1}}\frac{\left(N_{rx}^{'}-\eta\right)\left(N_{tx}^{'}-\eta\right)}{n_{\min}-\eta}.\label{eq:31}
\end{equation}
\end{thm}
\begin{IEEEproof}
By using SVD, we can decompose $\boldsymbol{H}=\boldsymbol{M}\boldsymbol{\Sigma}\boldsymbol{N}$
where $\boldsymbol{M}$ is the left singular matrix of $\boldsymbol{H}$,
$\boldsymbol{N}$ is the right singular matrix of $\boldsymbol{H}$,
$\boldsymbol{\Sigma}={\rm diag}\left(\Sigma_{1},\dots,\Sigma_{N_{\min}}\right)$
is a diagonal matrix whose $l$-th entry is the singular value of
$\boldsymbol{H}$ and $\Sigma_{1}\geq\cdots\geq\Sigma_{N_{\min}}$.
According to the Cauchy's Interlacing theorem \cite{hwang2004cauchy},
it follows that $\Sigma_{1}\geq\bar{\Sigma}_{1}\geq\cdots\geq\bar{\Sigma}_{n_{\min}}\geq\Sigma_{n_{\min}}\geq\cdots\geq\Sigma_{N_{\min}}$.
Therefore, the rate of MIMO-FAS can be upper bounded by 
\begin{equation}
R\left({\rm SNR}\right)=\sum_{l=1}^{n_{\min}}\log\left(1+\tilde{p}_{l}^{*}\Sigma_{l}^{2}\right),\label{eq:33}
\end{equation}
where $\tilde{p}_{l}^{*}=\left[\mu-\frac{1}{\Sigma_{l}^{2}}\right]_{0}^{+}$
and ${\rm SNR}=\sum\tilde{p}_{l}^{*}$. At high SNR, (\ref{eq:33})
can be simplified as 
\begin{equation}
R\left({\rm SNR}\right)=\sum_{l=1}^{n_{\min}}\log\left(1+\frac{{\rm SNR}}{n_{\min}}\Sigma_{l}^{2}\right),\label{eq:34}
\end{equation}
since the rate of using equal power allocation approaches to that
of waterfilling power allocation as ${\rm SNR}$ increases \cite{5733445}.
Consequently, the outage probability of MIMO-FAS can be lower bounded
by 
\begin{equation}
P_{{\rm out}}\left({\rm SNR},r\right)=\mathbb{{P}}\left\{ R\left({\rm SNR}\right)<r\log{\rm SNR}\right\} .\label{eq:35}
\end{equation}

At high SNR, the outage probability is rewritten as \cite{1197843}
\begin{equation}
P_{{\rm out}}\left({\rm SNR},r\right)=\mathbb{{P}}\left\{ \sum_{l=1}^{n_{\min}}\log\left(1+{\rm SNR}\Sigma_{l}^{2}\right)<r\log{\rm SNR}\right\} ,\label{eq:36}
\end{equation}
since 
\begin{align}
 & \lim_{{\rm SNR}\rightarrow\infty}\frac{\log\left(\mathbb{{P}}\left\{ \sum_{l=1}^{n_{\min}}\log\left(1+\frac{{\rm SNR}}{n_{\min}}\Sigma_{l}^{2}\right)<r\log{\rm SNR}\right\} \right)}{\log{\rm SNR}}\nonumber \\
= & \lim_{{\rm SNR}\rightarrow\infty}\frac{\log\left(\mathbb{{P}}\left\{ \sum_{l=1}^{n_{\min}}\log\left(1+\frac{{\rm SNR}}{n_{\min}}\Sigma_{l}^{2}\right)<r\log{\rm SNR}\right\} \right)}{\log\frac{{\rm SNR}}{n_{\min}}}\nonumber \\
= & \lim_{{\rm SNR}\rightarrow\infty}\frac{\log\left(\mathbb{{P}}\left\{ \sum_{l=1}^{n_{\min}}\log\left(1+{\rm SNR}\Sigma_{l}^{2}\right)<r\log{\rm SNR}\right\} \right)}{\log{\rm SNR}}.\label{eq:39}
\end{align}

Using (\ref{eq:24}), we can obtain the outer bound on the diversity
gain for a fixed multiplexing gain $r$ as 
\begin{equation}
P_{{\rm out}}\left({\rm SNR},r\right)\dot{=}~\mathbb{{P}}\left\{ \sum_{l=1}^{n_{\min}}\log\left(1+{\rm SNR}\Sigma_{l}^{2}\right)<r\log{\rm SNR}\right\} .\label{eq:40}
\end{equation}
According to \cite{6185738}, the joint probability density function
(PDF) of $\boldsymbol{\Sigma}^{2}$ is given by 
\begin{align}
f\left(\boldsymbol{\Sigma}^{2}\right) & =\sum_{\boldsymbol{q}}\frac{\left(-1\right)^{\frac{N_{\min}\left(N_{\min}-1\right)}{2}}\mathcal{A}}{N_{\min}!\triangle\left(\boldsymbol{q}\right)}\triangle\left(\boldsymbol{\Sigma}^{2}\right)\times\nonumber \\
 & \qquad\qquad{\rm det}\ensuremath{\left(\boldsymbol{\Sigma}_{k}^{2\left(q_{l}+N_{\max}-N_{\min}\right)}\right)},\label{eq:41}
\end{align}
where $\boldsymbol{q}=\left[q_{1},\dots,q_{N_{\max}}\right]^{T}$,
$N_{\min}=\min\left\{ N_{tx},N_{rx}\right\} $, 
\begin{align}
\mathcal{{A}} & =\frac{\prod_{k=1}^{N_{\min}}\lambda_{k}^{s}\prod_{l=1}^{N_{\max}}\lambda_{l}^{s'}{\rm det}\left(\left(-\lambda_{k}^{s}\right)^{q_{l}}\right)}{\triangle\left(\boldsymbol{\lambda}^{s}\right)\triangle(\boldsymbol{\lambda}^{s'})}\times\nonumber \\
 & \frac{{\rm det}\ensuremath{\left(\left.\left(\lambda_{k}^{s'}\right)^{q_{l}+N_{\max}-N_{\min}}\right|_{k=1}^{N_{\min}},\left.\left(\lambda_{i}^{s'}\right)^{N_{\max}-k}\right|_{k=N_{\min}+1}^{N_{\max}}\right)}}{\prod_{l=1}^{N_{\min}}\left(q_{l}+N_{\max}-N_{\min}\right)!},\label{eq:42}
\end{align}
where 
\begin{equation}
\left\{ \begin{aligned}s & =\left\{ s|N_{s}=\min\left\{ N_{tx},N_{rx}\right\} \right\} ,\\
s' & =\left\{ s'|N_{s'}=\max\left\{ N_{tx},N_{rx}\right\} \right\} ,\\
\boldsymbol{\lambda}^{s} & =\left[\lambda_{1}^{s},\dots,\lambda_{N_{s}}^{s}\right]^{T},
\end{aligned}
\right.
\end{equation}
$\triangle\left(\boldsymbol{\lambda}^{s}\right)$ denotes the Vandermonde
determinant of vector $\boldsymbol{\lambda}^{s}$, ${\rm det}\left(f\left(k,l\right)\right)$
is the determinant of a matrix with the $\left(k,l\right)$-th entry
given by the function $f\left(k,l\right)$ and $q_{l}=b_{l}+N_{\max}-l$.
In addition, $\boldsymbol{b}=\left[b_{1},\dots,b_{N_{\max}}\right]^{T}$
is the irreducible representation of unitary group such that $b_{1}\geq\cdots\geq b_{N_{\max}}\geq0$
are integers.

Conventionally, we can analyze the outer bound by simplifying the
joint PDF of $\boldsymbol{\Sigma}^{2}$ and then analyzing the exponents
of $\Sigma_{l}$. Nevertheless, it is found that no simplification
can be made to keep the exponents of $\Sigma_{l}$ tractable when
the rows and columns of $\boldsymbol{H}$ are fully correlated \cite{4016317}.
To alleviate this problem, we employ Lemma \ref{lem:lemma1}, Lemma
\ref{lem:lemma2}, and Lemma \ref{lem:lemma3}.

Specifically, according to \cite{gallager2008principles}, the PDF
of $\boldsymbol{H}$ can be obtained by removing the dependent entries
and the PDF of the singular values of $\boldsymbol{H}$ can be obtained
via coordinate changes \cite{978730}. Using Lemma \ref{lem:lemma1},
we know that $\boldsymbol{J}_{s}$ can be represented by a full rank
symmetric finite-size matrix, i.e, $\boldsymbol{J}_{{\rm red}}^{s}\in\mathbb{{R}}^{N_{s}^{'}\times N_{s}^{'}}$
and ${\rm rank}\left(\boldsymbol{J}_{{\rm red}}^{s}\right)=N_{s}^{'}$.
It follows that $\boldsymbol{H}$ and $\boldsymbol{G}$ in (\ref{eq:3})
can be rewritten as $N_{rx}^{'}\times N_{tx}^{'}$ matrices with the
same PDFs. From Lemma \ref{lem:lemma2}, it is known that the DMT
of any MIMO system with channel $\boldsymbol{H}\in\mathbb{C}^{N_{rx}^{'}\times N_{tx}^{'}}$
is the same as that of a system with channel $\boldsymbol{G}\in\mathbb{C}^{N_{rx}^{'}\times N_{tx}^{'}}$.
Since $\boldsymbol{G}$ is an independent and identically distributed
(i.i.d.)~circularly symmetric complex Gaussian matrix, it follows
that ${\rm rank}\left(\boldsymbol{G}\right)=N_{\min}^{'}=\min\{N_{rx}^{'},N_{tx}^{'}\}$
with probability one. Using Lemma \ref{lem:lemma3}, we can conclude
that the outer bound of the DMT of MIMO-FAS is a piecewise linear
function connecting the points as given in (\ref{eq:30}) and (\ref{eq:31}). 
\end{IEEEproof}
Using Theorem \ref{thm:thm4}, we can now derive the optimal DMT of
MIMO-FAS by considering its outer bound and inner bound. Specifically,
the DMT of QR MIMO-FAS can be regarded as the inner bound of the DMT
of MIMO-FAS.
\begin{thm}
\label{thm:thm5} The DMT of QR MIMO-FAS is equivalent to the outer
bound of the DMT of MIMO-FAS, and thus it is also the optimal DMT
of MIMO-FAS. 
\end{thm}
\begin{IEEEproof}
At high SNR, (\ref{eq:21}) can be simplified as 
\begin{equation}
R_{{\rm QR}}\left({\rm SNR}\right)=\sum_{l=1}^{n_{\min}}\log\left(1+\frac{{\rm SNR}}{n_{tx}}\bar{\Sigma}_{l}^{2}\right)\label{eq:43}
\end{equation}
since the rate of using equal power allocation approaches to that
of waterfilling power allocation as ${\rm SNR}$ increases \cite{5733445}.
Consequently, the outage probability is characterized by 
\begin{equation}
P_{{\rm QR}}^{{\rm out}}\left({\rm SNR},r\right)=\mathbb{{P}}\left\{ R_{{\rm QR}}\left({\rm SNR}\right)<r\log{\rm SNR}\right\} .\label{eq:44}
\end{equation}
At high SNR, the difference between (\ref{eq:34}) and (\ref{eq:43})
is 
\begin{equation}
R\left({\rm SNR}\right)-R_{{\rm QR}}\left({\rm SNR}\right)\approx\sum_{l=1}^{n_{\min}}\log\left(\frac{n_{tx}\Sigma_{l}^{2}}{n_{\min}\bar{\Sigma}_{l}^{2}}\right)=c_{0},\label{eq:45}
\end{equation}
which is a constant.\footnote{Note that a constant gap amounts to a finite scaling of SNR. This
implies that $R\left({\rm SNR}\right)=R_{{\rm QR}}\left({\rm SNR}+\gamma\right)$
for some $\gamma>0$.} As $\lim_{c\rightarrow\infty}\log\left(\frac{1+c}{c}\right)=0$,
the approximation of (\ref{eq:45}) is tight as ${\rm SNR}\rightarrow\infty$.
As such, we have 
\begin{align}
\lim_{{\rm SNR}\rightarrow\infty}\frac{R_{{\rm QR}}\left({\rm SNR}\right)}{\log{\rm SNR}} & =\lim_{{\rm SNR}\rightarrow\infty}\frac{R\left({\rm SNR}\right)-c_{0}}{\log{\rm SNR}}\nonumber \\
 & =\lim_{{\rm SNR}\rightarrow\infty}\frac{R\left({\rm SNR}\right)}{\log{\rm SNR}},\label{eq:47}
\end{align}
and 
\begin{align}
 & \lim_{{\rm SNR}\rightarrow\infty}\frac{\log\left(P_{{\rm QR}}^{{\rm out}}\left({\rm SNR},r\right)\right)}{\log{\rm SNR}}\nonumber \\
= & \lim_{{\rm SNR}\rightarrow\infty}\frac{\log\left(\mathbb{P}\left\{ R\left({\rm SNR}\right)<r\log{\rm SNR}+c_{0}\right\} \right)}{\log{\rm SNR}}\nonumber \\
= & \lim_{{\rm SNR}\rightarrow\infty}\frac{\log\left(\mathbb{P}\left\{ R\left({\rm SNR}\right)<r\log{\rm SNR}\right\} \right)}{\log{\rm SNR}}\nonumber \\
= & \lim_{{\rm SNR}\rightarrow\infty}\frac{\log\left(P_{{\rm out}}\left({\rm SNR},r\right)\right)}{\log{\rm SNR}}.\label{eq:51}
\end{align}
Thus, the DMT of QR MIMO-FAS is equivalent to the outer bound, and
also the optimal DMT of MIMO-FAS. 
\end{IEEEproof}
It is challenging to explicitly obtain $N_{s}^{'}$ because $\boldsymbol{J}_{s}$
might be near to being singular. It is also unclear how $\boldsymbol{J}_{s}$
can be reduced to/reconstructed from $\boldsymbol{J}_{{\rm red}}^{s}$.
To address these issues, we propose methods to reliably estimate $N_{s}^{'}$,
and to linearly transform $\boldsymbol{J}_{s}$ to $\boldsymbol{J}_{{\rm red}}^{s}$
and vice versa with a proof of certificates. These methods are based
on the following theorems.
\begin{thm}
\label{thm:thm6} Since $\boldsymbol{J}_{s}$ can be well-approximated
by $\hat{\boldsymbol{J}}_{s}$ where $\hat{\boldsymbol{J}}_{s}=\boldsymbol{U}_{s}\hat{\boldsymbol{\Lambda}}\boldsymbol{U}_{s}^{H}$
and $\hat{\boldsymbol{\Lambda}}={\rm diag}(\lambda_{1}^{s},\dots,\lambda_{N_{s}^{'}}^{s},0,\dots,0)$,
the rank of $\boldsymbol{J}_{s}$ can be estimated as $N_{s}^{'}$. 
\end{thm}
\begin{IEEEproof}
Define an arbitrarily small $\xi>0$ as the threshold where the numerical
values of the eigenvalues are negligible. According to \cite{hansen1987truncated},
$\boldsymbol{J}_{s}$ can be regularized by $\hat{\boldsymbol{J}}_{s}$,
where $\hat{\boldsymbol{J}}_{s}=\boldsymbol{U}_{s}\hat{\boldsymbol{\Lambda}}\boldsymbol{U}_{s}^{H}$,
$\hat{\boldsymbol{\Lambda}}={\rm diag}(\lambda_{1}^{s},\dots,\lambda_{N_{s}^{'}}^{s},0,\dots,0)$
and $\lambda_{l}^{s}<\xi$, $l\in\{N_{s}^{'}+1,\dots,N_{s}\}$. The
Frobenius norm between $\boldsymbol{J}_{s}$ and $\hat{\boldsymbol{J}}_{s}$
is given by 
\begin{equation}
E=\|\boldsymbol{J}_{s}-\hat{\boldsymbol{J}}_{s}\|_{F}=\sum_{l=N_{s}^{'}+1}^{N_{s}}\lambda_{l}^{s},\label{eq:52}
\end{equation}
which can be upper bounded by $(N_{s}-N_{s}^{'})\xi$. In practice,
we have $\lambda_{1}^{s}\gg\lambda_{N_{s}^{'}+1}^{s}>\ldots>\lambda_{N_{s}}^{s}$
and thus (\ref{eq:52}) is usually very small. Therefore, the rank
of $\boldsymbol{J}_{s}$ can be estimated as $N_{s}^{'}$. 
\end{IEEEproof}
\begin{thm}
\label{thm:thm7} For arbitrary $N_{s}=N_{1}^{s}\times N_{2}^{s}$
and a finite $W_{s}=W_{1}^{s}\times W_{2}^{s}$, $\boldsymbol{J}_{s}$
in (\ref{eq:2}) can be reduced to $\boldsymbol{J}_{{\rm red}}^{s}$
using the set $\mathcal{V}_{s}=\{\boldsymbol{v}_{l}^{s}|\boldsymbol{J}_{s}^{\left(l\right)}\tilde{\boldsymbol{v}}_{l}^{s}=\boldsymbol{0},l=N_{s}^{'}+1,\ldots,N_{s}\}$
where $\tilde{\boldsymbol{v}}_{l}^{s}=[\boldsymbol{v}_{l}^{s}~-1]^{T}$.
Conversely, $\boldsymbol{J}_{s}$ in (\ref{eq:2}) can be reconstructed
from $\boldsymbol{J}_{{\rm red}}^{s}$ with the set $\mathcal{V}_{s}$. 
\end{thm}
\begin{IEEEproof}
Let us introduce a vector $\tilde{\boldsymbol{v}}_{l}^{s}=[\boldsymbol{v}_{l}^{s}~-1]^{T}$
where $\boldsymbol{v}_{l}^{s}\in\mathbb{R}^{l}$ and $\boldsymbol{v}_{l}^{s}\neq\boldsymbol{0}$.
In addition, let us denote 
\begin{equation}
\boldsymbol{J}_{s}^{\left(l\right)}=\left[\begin{array}{cc}
\boldsymbol{J}_{{\rm sub}}^{\left(l-1\right)} & \boldsymbol{j}_{l}\\
\boldsymbol{j}_{l}^{T} & j_{l,l}
\end{array}\right].\label{eq:53}
\end{equation}
Note that $\boldsymbol{J}_{s}=\boldsymbol{J}_{s}^{\left(N_{s}\right)}$
and $\boldsymbol{J}_{{\rm red}}^{s}=\boldsymbol{J}_{s}^{(N_{s}^{'})}$.
If $N_{s}^{'}=N_{s}$, then $\mathcal{V}_{s}$ is an empty set and
$\boldsymbol{J}_{s}=\boldsymbol{J}_{{\rm red}}^{s}$. Thus, we can
focus on the case where $N_{s}^{'}<N_{s}$. Without loss of generality,
we may assume that the last column of $\boldsymbol{J}_{s}^{\left(l\right)}$
is a linear combination of the first $\left(l-1\right)$ columns of
$\boldsymbol{J}_{s}^{\left(l\right)}$. This implies that $\boldsymbol{J}_{s}^{\left(l\right)}\tilde{\boldsymbol{v}}_{l}^{s}=\boldsymbol{0}$,
i.e., $\tilde{\boldsymbol{v}}_{l}^{s}$ is in the null space of $\boldsymbol{J}_{s}^{\left(l\right)}$
and $\tilde{\boldsymbol{v}}_{l}^{s}\neq0$. If $\boldsymbol{J}_{s}^{\left(l\right)}\tilde{\boldsymbol{v}}_{l}^{s}=\boldsymbol{0}$
and $l\neq N_{s}^{'}$, we can reduce $\boldsymbol{J}_{s}^{\left(l\right)}$
to $\boldsymbol{J}_{{\rm sub}}^{\left(l-1\right)}$ by setting $\boldsymbol{v}_{l}^{s}\in\mathcal{V}_{s}$.
Otherwise, we have $\boldsymbol{J}_{s}^{\left(l\right)}=\boldsymbol{J}_{{\rm red}}^{s}$
since $\boldsymbol{J}_{s}^{\left(l\right)}$ must be a full rank matrix.
Conversely, $\boldsymbol{J}_{s}^{\left(l\right)}$ can be reconstructed
from $\boldsymbol{J}_{{\rm sub}}^{\left(l-1\right)}$ if $\boldsymbol{v}_{l}^{s}$
is given. Specifically, we can define $\boldsymbol{j}_{l}\triangleq\boldsymbol{J}_{{\rm sub}}^{\left(l-1\right)}\boldsymbol{v}_{l}^{s}$
and $j_{l,l}\triangleq\boldsymbol{j}_{l}^{T}\boldsymbol{v}_{l}^{s}$
and they are sufficient to reconstruct $\boldsymbol{J}_{s}^{\left(l\right)}$.
Hence, $\boldsymbol{v}_{l}^{s}$ can be interpreted as a certificate
that $\boldsymbol{J}_{s}^{\left(l\right)}$ can be reduced to/reconstructed
from $\boldsymbol{J}_{{\rm sub}}^{\left(l-1\right)}$ for $l=\{N_{s}^{'}+1,\dots,N_{s}\}$. 
\end{IEEEproof}
\begin{table}
\caption{\label{tab:I} Estimation of $N_{s}^{'}$ for different $W_{s}$ in
terms of $\lambda^{2}$, where $\xi=10^{-3}$, $N_{s}=100$, $N_{1}^{s}=N_{2}^{s}$,
and $W_{1}^{s}=W_{2}^{s}$.}

\begin{centering}
\begin{tabular}{c|c|c|c|c|c|c}
\hline 
$W_{s}$  & $0.5\times0.5$  & $1\times1$  & $1.5\times1.5$  & $2\times2$  & $2.5\times2.5$  & $3\times3$\tabularnewline
\hline 
\hline 
$N_{s}^{'}$  & $13$  & $23$  & $34$  & $48$  & $60$  & $73$\tabularnewline
\hline 
$E$  & $0.001$  & $0.002$  & $0.003$  & $0.002$  & $0.005$  & $0.005$\tabularnewline
\hline 
\end{tabular}
\par\end{centering}
\begin{centering}
\vspace{0.2cm}
 
\par\end{centering}
{[}$^{\ddag}${]} Note that the figures regarding the diversity order
of MIMO-FAS described in \cite{WongIL} were based on an earlier version
of this paper considering the spatial correlation in 2D environments
only. The correct diversity orders for MIMO-FAS in 3D environments
should be referred to this table. For example, for MIMO-FAS with $0.5\lambda\times0.5\lambda$
FAS at both ends, the diversity order is $13\times13=169$ which is
much higher than originally reported. 
\end{table}

The proposed methods enable us to estimate $N_{s}^{'}$ for given
$N_{1}^{s}$, $N_{2}^{s}$, $W_{1}^{s}$, and $W_{2}^{s}$. Furthermore,
they allow us to verify that $\boldsymbol{J}_{s}$ indeed can be reduced
to (or reconstructed from) $\boldsymbol{J}_{{\rm red}}^{s}$ with
proof of certificates. An example of the estimations of $N_{s}^{'}$
is given in Table \ref{tab:I}. This table can help us better understand
the performance of MIMO-FAS. For example, by substituting the estimation
of $N_{s}^{'}$ into Theorem \ref{thm:thm4}, we observe that MIMO-FAS
yields massive diversity gains if $r<n_{\min}$. Thus, it is worth
investigating how we can leverage MIMO-FAS effectively. To answer
this question, we introduce the $q$-outage capacity.
\begin{defn}
\label{def:def8} The $q$-outage capacity of a system is defined
as 
\begin{equation}
C_{{\rm sys}}^{q}=q\left(1-\bar{P}_{{\rm sys}}^{{\rm out}}\left({\rm SNR},q\right)\right),\label{eq:54}
\end{equation}
where 
\begin{equation}
\bar{P}_{{\rm sys}}^{{\rm out}}\left({\rm SNR},q\right)=\mathbb{{P}}\left\{ R_{{\rm sys}}\left({\rm SNR}\right)<q\right\} ,\label{eq:55}
\end{equation}
where $\bar{P}_{{\rm sys}}^{{\rm out}}\left(q\right)$ is the outage
probability of a system for a fixed target rate or transmission rate
$q$ independent of the SNR. It is worth noting that (\ref{eq:54})
differs from the $\epsilon$-outage capacity as it is not measuring
the largest $q$ such that $\bar{P}_{{\rm sys}}^{{\rm out}}\left({\rm SNR},q\right)\leq\epsilon$.
Instead, (\ref{eq:54}) is interpreted as the average rate that a
system can reliably transmit over period of time such that the statistics
of the fading do not change. Furthermore, both $q$ and $\bar{P}_{{\rm sys}}^{{\rm out}}\left({\rm SNR},q\right)$
play important roles in (\ref{eq:54}). For example, if $q\approx0$,
we usually have $\bar{P}_{{\rm sys}}^{{\rm out}}\left({\rm SNR},q\right)\approx0$.
If $q$ is large, then we typically have $\bar{P}_{{\rm sys}}^{{\rm out}}\left({\rm SNR},q\right)\approx1$.
Nevertheless, both cases yield $C_{{\rm sys}}^{q}\approx0$. In between
the two extremes, we have $C_{{\rm sys}}^{q}\gg0$ and there is an
optimal $q$ for a system such that (\ref{eq:54}) is maximized. In
addition, the $q$-outage capacity gain of MIMO-FAS over a traditional
antenna system can be characterized as 
\begin{multline}
C_{{\rm MIMO\text{-}FAS}}^{q}-C_{{\rm sys}}^{q}\\
=q\left[\bar{P}_{{\rm {\rm sys}}}^{{\rm out}}\left({\rm SNR},q\right)-\bar{P}_{{\rm MIMO\text{-}FAS}}^{{\rm out}}\left({\rm SNR},q\right)\right].\label{eq:56}
\end{multline}
Using (\ref{eq:56}), we can easily see the benefits that can be harnessed
by MIMO-FAS over a traditional antenna system. 
\end{defn}

\section{Results and Discussions}

Here, we present the analytical and Monte Carlo simulation results
to evaluate the performance of MIMO-FAS. For brevity, we focus on
a symmetric MIMO-FAS design where $N_{1}^{s}=N_{2}^{s}$, $W_{1}^{s}=W_{2}^{s}$
and $W_{rx}=W_{tx}$. Unless stated otherwise, we assume that $N_{s}=100$,
$n_{s}=4$, $W_{s}=1\lambda^{2}$, $\delta_{s}^{2}=1$ and ${\rm SNR}=30{\rm dB}$.
We also consider multiple schemes based on different combinations
of techniques to highlight the respective gains and effects. These
benchmarking schemes are:\footnote{Note that \cite{ma2022mimo} has proposed a solution where the antennas
can move to locally optimal positions. We can interpret these positions
as activating some ports as $N_{s}\rightarrow\infty$. However, the
solution cannot be employed here due to two impediments. Firstly,
the spatial correlation of these positions cannot be obtained. Secondly,
we are considering the cases where the positions might be discrete.
Therefore, the solution is not considered in this paper.} 
\begin{itemize}
\item Optimal MIMO-FAS: It considers the MIMO-FAS setup that utilizes an
exhaustive search, SVD and waterfilling for port selection, beamforming
and power allocation, respectively. 
\item QR MIMO-FAS: This is the proposed MIMO-FAS that employs strong RRQR
factorization, SVD and waterfilling power allocation for a suboptimal
solution. 
\item Greedy MIMO-FAS: This is the MIMO-FAS that employs greedy selection,
SVD and waterfilling power allocation for an efficient solution in
the low SNR regime. 
\item Random MIMO-FAS: It randomly activates the ports and uses SVD as well
as waterfilling power allocation. 
\item MIMO: This refers to the traditional MIMO that employs SVD and waterfilling
power allocation. Unless otherwise stated, we assume that the number
of antennas is $n_{s}^{{\rm mimo}}=4$ and the antennas are spatially
correlated based on (\ref{eq:1}). 
\item MIMO-AS: This refers to the traditional MIMO antenna selection system
that employs strong RRQR factorization, SVD and waterfilling power
allocation. Unless stated otherwise, the number of active antennas
is $n_{s}^{{\rm mimo-as}}=4$. Also, a maximum number of antennas
is considered in the given surface where the antennas are placed based
on a grid structure with at least half a wavelength apart, and they
are spatially correlated due to (\ref{eq:1}). 
\end{itemize}
\begin{figure}
\centering{}\includegraphics[scale=0.6]{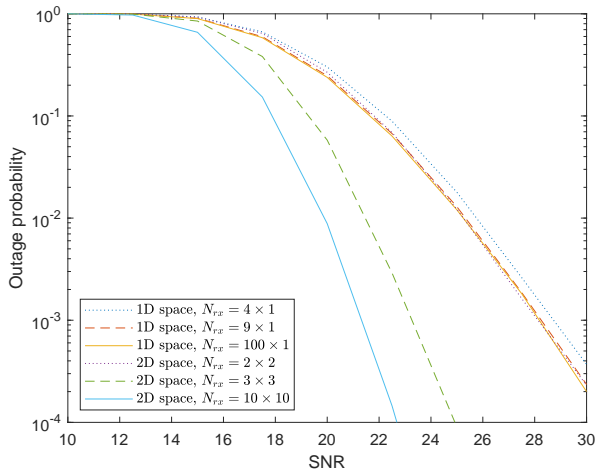}\caption{Outage probability of FAS versus SNR for different number of ports
and dimensional space, with $q=7{\rm bps/Hz}$.}
\label{fig:OPvsSNR} 
\end{figure}

\begin{figure}
\begin{centering}
\subfloat[]{\begin{centering}
\includegraphics[scale=0.6]{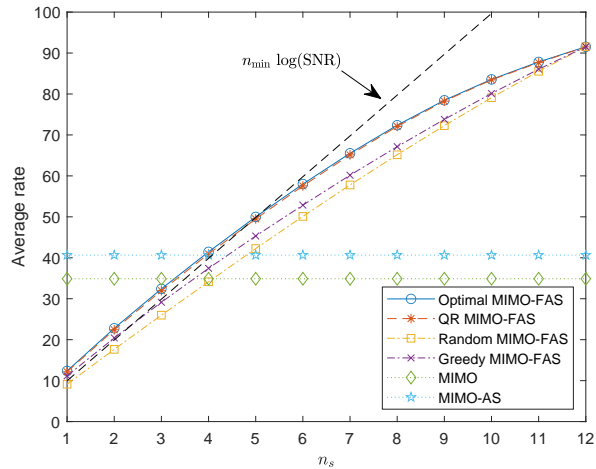} 
\par\end{centering}
}
\par\end{centering}
\begin{centering}
\subfloat[]{\begin{centering}
\includegraphics[scale=0.6]{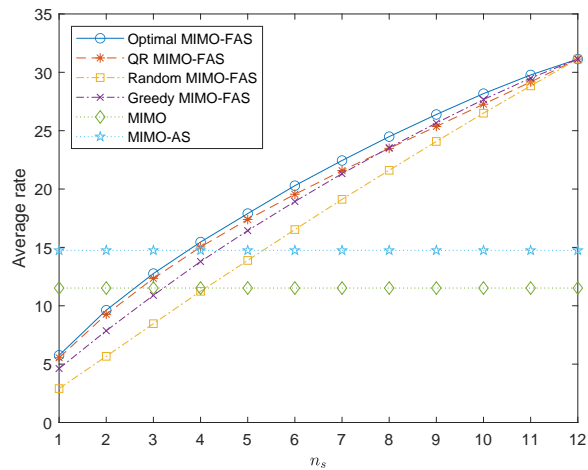} 
\par\end{centering}
}
\par\end{centering}
\centering{}\subfloat[]{\begin{centering}
\includegraphics[scale=0.6]{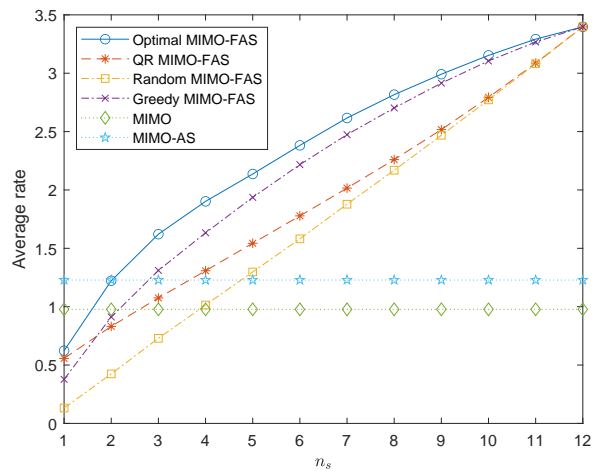} 
\par\end{centering}
}\caption{Average rates of the benchmarking schemes for different values of
$n_{s}$: a) ${\rm {SNR}=30}$dB; b) ${\rm {SNR}=10}$dB; c) ${\rm {SNR}=-10}$dB.}
\label{fig:Rvsns} 
\end{figure}

To highlight the superiority of 2D space, we first consider a simplified
scenario where there is only a single fluid antenna at the receiver.
Fig. \ref{fig:OPvsSNR} shows the outage probability of FAS versus
SNR for various number of ports and dimensional space. Here, the outage
probability is obtained using (\ref{eq:55}) by replacing $R_{{\rm sys}}\left({\rm SNR}\right)$
with $R_{{\rm QR}}\left({\rm SNR}\right)$. Given the same number
of ports, the ports that are distributed in 2D space can achieve a
much lower outage probability as compared to the ones that are distributed
in 1D space. This improvement can be explained from the fact that
a 2D space has an additional dimension for the fluid antenna to move
around. Hence, it contains more spatial diversity and yields a lower
outage probability. This suggests that the ports in MIMO-FAS should
be designed using the entire 2D space for better performance.

Next in Fig. \ref{fig:Rvsns}, we study the average rates of the benchmarking
schemes for different $n_{s}$ at different SNR. Since the optimal
MIMO-FAS is computed using an exhaustive search, we set $N_{s}=12$
where $N_{1}^{s}=3$ and $N_{2}^{s}=4$. Fig. \ref{fig:Rvsns}(a)
illustrates that QR MIMO-FAS achieves a similar average rate as compared
to the optimal MIMO-FAS at high SNR. Furthermore, the average rate
of QR MIMO-FAS scales like $n_{s}\log{\rm SNR}$ when $n_{s}$ ranges
from $1$ to $6$. Nevertheless, it suffers from a diminishing rate
gain when $n_{s}$ ranges from $7$ to $12$. Thus, MIMO-FAS is most
effective when each active port has the freedom of being at least
half a wavelength apart from each other. In addition, the average
rate of QR MIMO-FAS outperforms traditional MIMO when the number of
active ports or antennas is the same (i.e., $n_{s}=4$). This is because
QR MIMO-FAS activates the optimal ports in each realization, which
reduces the spatial correlation effect. \textcolor{black}{In Fig.
\ref{fig:Rvsns}(b), we further observe that QR MIMO-FAS provides
a higher sum-rate in the medium SNR regime when $n_{s}$ is small
while greedy MIMO-FAS yields a better performance when $n_{s}$ is
large.} Nevertheless, as shown in Fig. \ref{fig:Rvsns}(c), greedy
MIMO-FAS is generally more efficient than QR MIMO-FAS in the low SNR
regime. \textcolor{black}{These results suggest that an efficient
scheme with low complexity is still required to maximize the rate
of MIMO-FAS in the medium and low SNR regimes.} Besides, QR MIMO-FAS
yields a similar or higher rate than MIMO-AS when the number of active
ports or antennas is the same.

\begin{figure}
\begin{centering}
\subfloat[]{\begin{centering}
\includegraphics[scale=0.6]{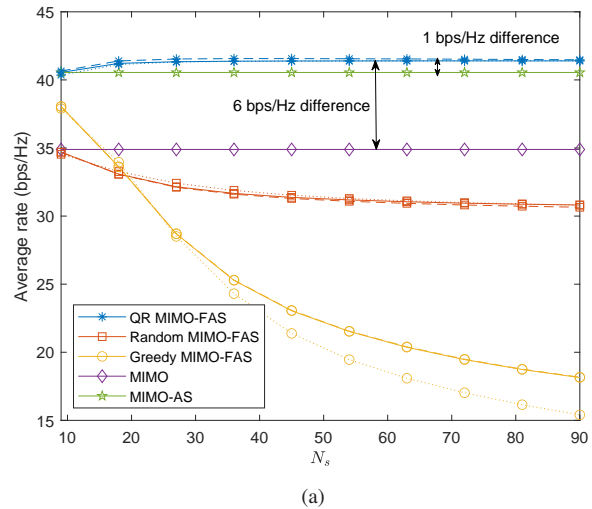} 
\par\end{centering}
}
\par\end{centering}
\centering{}\subfloat[]{\begin{centering}
\includegraphics[scale=0.6]{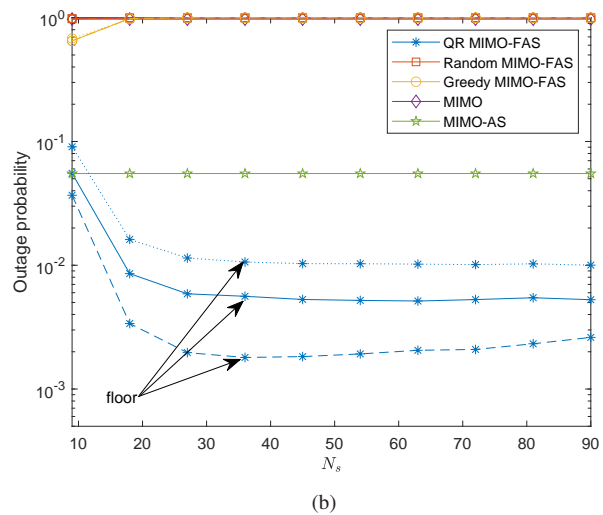} 
\par\end{centering}
}\caption{The performance of the benchmarking schemes for different values of
$N_{s}$: (a) average rate; (b) outage probability, with $q=39{\rm bps/Hz}$;
(solid) without mutual coupling; (dotted) liquid-based fluid antenna
with mutual coupling; (dashed) RF pixel-based fluid antenna with mutual
coupling.}
\label{fig:Pvsns} 
\end{figure}

Fig. \ref{fig:Pvsns} presents the average rates and outage probabilities
of the benchmarking schemes for different values of $N_{s}$. In these
results, we omit the optimal MIMO-FAS because it is difficult to perform
exhaustive search online for large $N_{s}$. \textcolor{black}{On
the other hand, since the activated ports can be placed very close
to each other, it would be useful to consider the mutual coupling
effect and investigate its effect on the performance of MIMO-FAS.
In particular, we consider two designs: liquid-based and RF pixel-based
fluid antennas. To make a fair comparison between the cases with and
without mutual coupling, we assume that $N_{1}^{s}=\left\lfloor \frac{W_{1}^{s}}{0.5}\right\rfloor +1$
and vary $N_{2}^{s}$ accordingly. This means the resolution in one
direction is fixed while we change the resolution of FAS in another
direction to examine the impact of mutual coupling. The details of
the mutual coupling model are given in Appendix III.}

\textcolor{black}{As seen in Fig. \ref{fig:Pvsns}, generally speaking,
the performance of QR MIMO-FAS and random MIMO-FAS with mutual coupling
are not vastly different from the case without mutual coupling regardless
of whether liquid-based or RF pixel-based fluid antenna is considered.
This suggests that the active ports can be placed close to each other,
typically much less than half a wavelength, and still yield a similar
performance. By contrast, the rate performance of greedy MIMO-FAS
appears to suffer more from mutual coupling as $N_{s}$ increases.
However, greedy MIMO-FAS is not supposed to work well here because
the setting is under the high SNR regime.}

\textcolor{black}{It is worth pointing out that for RF pixel-based
fluid antenna, mutual coupling can exist regardless of whether the
pixels are on or off. It is therefore essential to improve the S-matrix
via antenna design in order to achieve a good performance. However,
the advantage of RF pixel-based fluid antenna is that the mutual coupling
matrix is deterministic given $N_{1}^{s}$ and $N_{2}^{s}$ regardless
of which pixels are the active ones. Thus, one can directly obtain
the optimal port selection, beamforming and power allocation while
taking into account of the mutual coupling effect. Moreover, matching
networks can be employed directly to further improve the performance
of MIMO-FAS but this technique is not considered in our results.}

\textcolor{black}{Based on the above observations, we hence focus
on the performance of MIMO-FAS without mutual coupling effect.} In
Fig. \ref{fig:Pvsns}(a), we see that the average rates of random
MIMO-FAS and greedy MIMO-FAS generally decrease as $N_{s}$ increases.
This suggests that efficient port selection in MIMO-FAS is essential.
Furthermore, the average rate of QR MIMO-FAS is $1{\rm bps/Hz}$ higher
than that of MIMO-AS and $6{\rm bps/Hz}$ higher than that of MIMO
when $N_{s}$ is large. In Fig. \ref{fig:Pvsns}(b), the outage probabilities
of random MIMO-FAS, Greedy MIMO-FAS and MIMO are near one (i.e., $0.99$)
while the outage probability of MIMO-AS is in the order of $10^{-2}$.
In contrast, the outage probability of QR MIMO-FAS is much lower (i.e.,
the order of $10^{-3}$). Nevertheless, the outage probability of
QR MIMO-FAS decreases to a floor as $N_{s}$ continues to increase.
This limitation is due to the fact that there are approximately $N_{s}^{'}$
diversity in $\boldsymbol{J}_{s}$ for a fixed $W_{s}$. Therefore,
the outage probability of QR MIMO-FAS is limited by $N_{s}^{'}$ for
a fixed $W_{s}$ (see, Table \ref{tab:I}).

\begin{figure}
\begin{centering}
\subfloat[]{\begin{centering}
\includegraphics[scale=0.6]{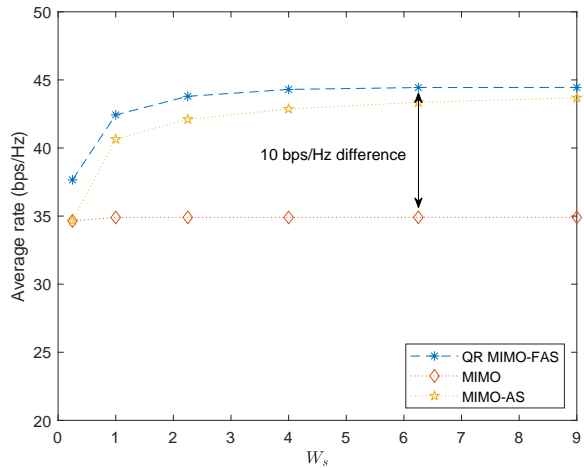} 
\par\end{centering}
}
\par\end{centering}
\centering{}\subfloat[]{\begin{centering}
\includegraphics[scale=0.6]{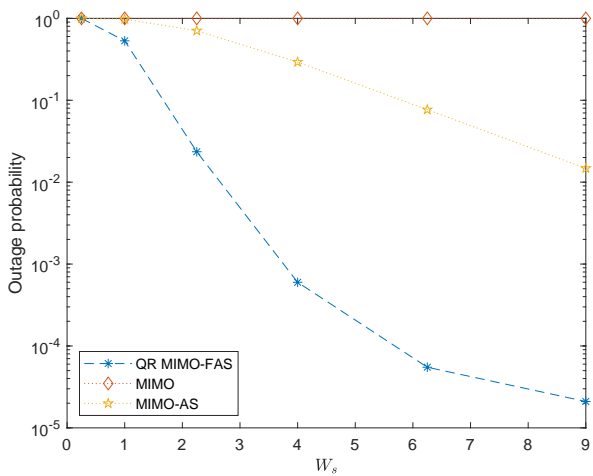} 
\par\end{centering}
}\caption{The performance of QR MIMO-FAS, MIMO and MIMO-AS for different values
of $W_{s}$: (a) average rate; (b) outage probability, with $q=42.5$bps/Hz.}
\label{fig:PvsW} 
\end{figure}

To verify this explanation, we further investigate the outage probabilities
of QR MIMO-FAS, MIMO and MIMO-AS for different values of $W_{s}$.
For brevity, we omit random MIMO-FAS and greedy MIMO-FAS as we now
know that these schemes do not provide an effective performance for
large $N_{s}$. As seen in Fig. \ref{fig:PvsW}(a), the average rates
of QR MIMO-FAS, MIMO and MIMO-AS increase and then plateau. Nevertheless,
as shown in Fig. \ref{fig:PvsW}(b), the outage probabilities of QR
MIMO-FAS decrease without bound as $W_{s}$ increases if $N_{s}$
is sufficiently large. By analyzing Table \ref{tab:I}, we can observe
that although $N_{s}$ remains fixed, $N_{s}^{'}$ generally increases
if $W_{s}$ is increased. In contrast, the outage probability of MIMO
is always near to being one since the diversity gain of MIMO remains
the same even if $W_{s}$ is increased. On the other hand, the outage
probability of MIMO-AS decreases at a slower rate due to the lower
resolution (i.e., the number of antennas is less than the number of
ports in the given space). Overall, Figs. \ref{fig:Pvsns} and \ref{fig:PvsW}
suggest that the value of $N_{s}$ be determined by $W_{s}$.

\begin{figure}
\begin{centering}
\includegraphics[scale=0.6]{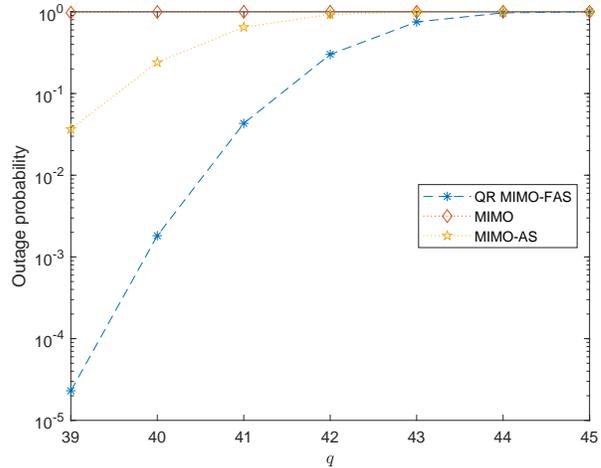} 
\par\end{centering}
\centering{}\caption{The outage probabilities of QR MIMO-FAS, MIMO and MIMO-AS for different
values of $q$.}
\label{fig:OPvsq} 
\end{figure}

From the above results, one might be amazed with the rate improvement
of QR MIMO-FAS. Nevertheless, we highlight that the superiority of
QR MIMO-FAS lies in the diversity gain. In particular, QR MIMO-FAS
can reduce its outage probability to a much lower value than MIMO
and MIMO-AS if $q$ is low (e.g., $q<n_{\min}\log{\rm SNR}$). To
examine this phenomenon more closely, Fig. \ref{fig:OPvsq} illustrates
the outage probabilities of QR MIMO-FAS, MIMO and MIMO-AS for different
$q$. Within $6{\rm bits/Hz}$, we see that the outage probability
of QR MIMO-FAS reduces at a steeper rate (e.g., from the order of
$10^{-1}$ to $10^{-5}$) while the outage probability of MIMO-AS
reduces at a slower rate (e.g., from the order of $10^{-1}$ to $10^{-2}$)
and the outage probability of MIMO remains roughly the same.

\begin{figure}
\begin{centering}
\includegraphics[scale=0.6]{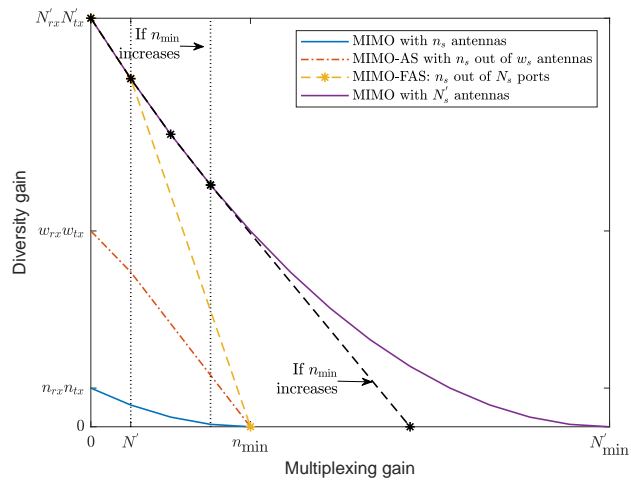} 
\par\end{centering}
\centering{}\caption{The optimal DMT of MIMO-FAS, MIMO and MIMO-AS.}
\label{fig:DMT} 
\end{figure}

To understand this at a more fundamental level, we present the DMT
of QR MIMO-FAS in Fig. \ref{fig:DMT}. Note that the DMT of QR MIMO-FAS
is also the optimal DMT of MIMO-FAS. As shown in Table \ref{tab:I},
the value of $N_{s}^{'}$ depends on $W_{s}$ as long as $N_{s}\geq N_{s}^{'}$.
In Fig. \ref{fig:DMT}, it can be seen that the diversity gain of
MIMO-FAS is much superior than that of an $n_{rx}\times n_{tx}$ MIMO
system for a fixed $r$. For example, the maximum diversity of $4\times4$
MIMO is $16$ as $r\rightarrow0$. This is because the optimal DMT
of a traditional $n_{rx}\times n_{tx}$ MIMO system is a piecewise
linear function connecting the point $\left(r,\left(n_{rx}-r\right)\left(n_{tx}-r\right)\right)$\cite{1197843}.
Meanwhile, the maximum diversity of MIMO-AS is limited by $w_{rx}w_{tx}$.
For instance, if $W_{s}=1\lambda^{2}$, the maximum diversity of MIMO-AS
is $81$. In contrast, the maximum diversity gain of MIMO-FAS is approximately
$23\times23=529\gg\left\{ 81,16\right\} $ if $W_{rx}=W_{tx}=1\lambda^{2}$.
Hence, the diversity gain of MIMO-FAS is massive because outage only
occurs when all the ports experience deep fades. To obtain the same
diversity gain at multiplexing gain $r$ from $0$ to $N'$, a traditional
$N_{rx}^{'}\times N_{tx}^{'}$ MIMO would have been required. However,
the downside is that it can only have at most $n_{\min}$ multiplexing
gain.

\begin{figure}
\begin{centering}
\includegraphics[scale=0.6]{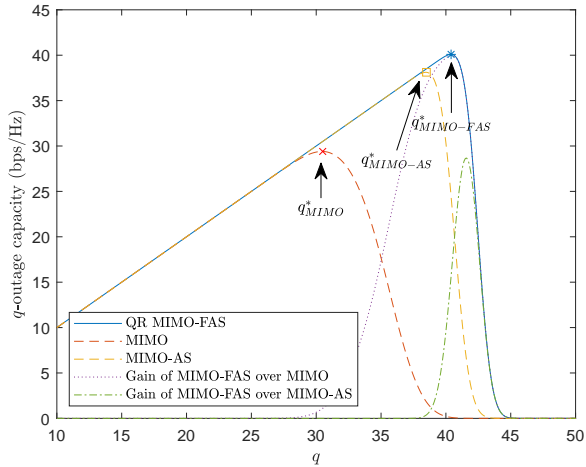} 
\par\end{centering}
\centering{}\caption{The $q$-outage capacity of MIMO-FAS, MIMO and MIMO-AS.}
\label{fig:qC} 
\end{figure}

Finally, we investigate the $q$-outage capacity to showcase the benefits
of MIMO-FAS. Fig. \ref{fig:qC} presents the $q$-outage capacities
of QR MIMO-FAS, MIMO and MIMO-AS as well as the $q$-outage capacity
gain of QR MIMO-FAS over MIMO and MIMO-AS. For ease of exposition,
the optimal $q^{*}$ of QR MIMO-FAS, MIMO and MIMO-AS are denoted
as $q_{{\rm MIMO\text{-}FAS}}^{*}$, $q_{{\rm MIMO}}^{*}$, $q_{{\rm MIMO\text{-}AS}}^{*}$,
respectively. As it is seen, the outage capacities of the schemes
increase up to $q^{*}$ and decrease thereafter. To the left side
of $q^{*}$, the capacity is limited by $q$ (i.e., rate) since $\bar{P}_{{\rm sys}}^{{\rm out}}\left({\rm SNR},q\right)$
is small. To the right side of $q^{*}$, the capacity is limited by
the outage probability because $\bar{P}_{{\rm sys}}^{{\rm out}}\left({\rm SNR},q\right)$
is large. Since MIMO provides limited diversity gain and MIMO-AS has
a limited number of antennas within a given space, both schemes fail
to achieve certain $q$-outage capacity achievable by QR MIMO-FAS.
This suggests that MIMO-FAS can reliably deliver a much higher rate
than traditional MIMO and MIMO-AS systems.

\section{Conclusions}

In this paper, we analyzed the performance limits of MIMO-FAS. To
this end, we developed a system model for MIMO-FAS where a 2D fluid
antenna surface was used at both ends, while taking into account of
the spatial correlation of the ports. We then proposed a suboptimal
scheme to maximize the rate of MIMO-FAS at high SNR via joint port
selection, beamforming and power allocation, namely QR MIMO-FAS. One
key contribution was the derivation of the outer bound of the DMT
for MIMO-FAS. Through the outer bound and the proposed scheme, we
then obtained the optimal DMT of MIMO-FAS which revealed the fundamental
limits of MIMO-FAS. Extensive results were presented, illustrating
that QR MIMO-FAS achieved a similar rate as compared to the optimal
MIMO-FAS in the high SNR regime. By fixing other MIMO-FAS parameters,
we found that the average rate and outage probability of QR MIMO-FAS
approached to a limit as $N_{s}$ increased. Likewise, the average
rate of QR MIMO-FAS improved up to a certain level as $W_{s}$ increased.
Nevertheless, the outage probability of QR MIMO-FAS decreased without
bound as $W_{s}$ increased. For the same multiplexing gain, we also
showed that MIMO-FAS achieved massive diversity gain as compared to
the traditional MIMO and MIMO-AS systems. Motivated by this, we further
illustrated that MIMO-FAS could reliably deliver a much higher rate
than the traditional MIMO and MIMO-AS systems in terms of $q$-outage
capacity that jointly considered both rate and outage probability.

\section*{Appendix I: Spatial Correlation of 2D Fluid Antenna Surface over
3D Scattering Environment}

Without loss of generality, let us refer to the position of the ($n_{1},n_{2}$)-port
as $\boldsymbol{n}_{l_{s}}=\left[0,\frac{n_{2}^{s}-1}{N_{2}^{s}-1}W_{2}^{s},\frac{n_{1}^{s}-1}{N_{1}^{s}-1}W_{1}^{s}\right]{}^{T}$
where ${\rm {map}}\left(n_{1},n_{2}\right)=l_{s}$ and $\lambda$
is the wavelength. Suppose a plane wave impinges on the fluid antenna
surface from azimuth angle $\varphi$ and elevation angle $\theta$.
Then, the array response vector can be expressed as \cite{Emil_1}
\begin{equation}
\boldsymbol{a}\left(\varphi,\theta\right)=\left[e^{j\frac{2\pi}{\lambda}\boldsymbol{k}\left(\varphi,\theta\right)^{T}\boldsymbol{n}_{1}\lambda},\dots,e^{j\frac{2\pi}{\lambda}\boldsymbol{k}\left(\varphi,\theta\right)^{T}\boldsymbol{n}_{N_{s}}\lambda}\right]^{T},\label{eq:A1}
\end{equation}
where 
\begin{equation}
\boldsymbol{k}\left(\varphi,\theta\right)=\left[\begin{array}{ccc}
\cos\left(\theta\right)\cos\left(\varphi\right) & \cos\left(\theta\right)\sin\left(\varphi\right) & \sin\left(\theta\right)\end{array}\right]^{T},\label{eq:A2}
\end{equation}
is the normalized wave vector.

Let us denote the spatial correlation matrix as $\boldsymbol{J}_{s}=\mathbb{{E}}\left\{ \boldsymbol{a}\left(\varphi,\theta\right)\boldsymbol{a}\left(\varphi,\theta\right)^{H}\right\} $.
From (\ref{eq:A1}), we know that the $\left(k_{s},l_{s}\right)$-th
entry of $\boldsymbol{J}_{s}$ can be expressed as 
\begin{align}
\left[\boldsymbol{J}_{s}\right]_{k_{s},l_{s}} & =\mathbb{{E}}\left\{ e^{j\frac{2\pi}{\lambda}\boldsymbol{k}\left(\varphi,\theta\right)^{T}\boldsymbol{n}_{k_{s}}\lambda}e^{-j\frac{2\pi}{\lambda}\boldsymbol{k}\left(\varphi,\theta\right)^{T}\boldsymbol{n}_{l_{s}}\lambda}\right\} \nonumber \\
 & =\mathbb{{E}}\left\{ e^{j2\pi\boldsymbol{k}\left(\varphi,\theta\right)^{T}\left(\boldsymbol{n}_{k_{s}}-\boldsymbol{n}_{l_{s}}\right)}\right\} ,\label{eq:A4}
\end{align}
where ${\rm {map}}\left(\tilde{n}_{1}^{s},\tilde{n}_{2}^{s}\right)=k_{s}$.
Using the Jacobi-Anger plane wave expansion \cite{Mehrem}, (\ref{eq:A4})
can be rewritten as 
\begin{multline}
\left[\boldsymbol{J}_{s}\right]_{k_{s},l_{s}}=4\pi\sum_{m=0}^{\infty}\sum_{n=-m}^{m}\left(-i\right)^{m}\times\\
\alpha_{m}^{n}Y_{m}^{n}\left(\frac{\boldsymbol{n}_{k_{s}}-\boldsymbol{n}_{l_{s}}}{\left\Vert \boldsymbol{n}_{k_{s}}-\boldsymbol{n}_{l_{s}}\right\Vert }\right)\times\\
j_{m}\left(2\pi\left\Vert \boldsymbol{n}_{k_{s}}-\boldsymbol{n}_{l_{s}}\right\Vert \right),\label{eq:A5}
\end{multline}
where $j_{m}\left(.\right)$ is the spherical Bessel function of the
first kind, $Y_{m}^{n}\left(\cdot\right)$ is the spherical harmonics,
and %Furthermore, in (\ref{eq:A5}), we have 
\begin{align}
\alpha_{m}^{n} & =\int_{\Omega}f_{\boldsymbol{k}\left(\varphi,\theta\right)}\left(\boldsymbol{k}\right)Y_{m}^{n}\left(\boldsymbol{k}\right)d\Omega\left(\boldsymbol{k}\right),\label{eq:A6}
\end{align}
where $\Omega\left(\boldsymbol{k}\right)$ is a surface element of
a unit sphere $\Omega$. For a 3D isotropic scattering environment,
we have $f_{\boldsymbol{k}\left(\varphi,\theta\right)}\left(\boldsymbol{k}\right)=\frac{1}{4\pi}$
and thus (\ref{eq:A6}) reduces to 
\begin{equation}
\alpha_{m}^{n}=\frac{1}{4\pi}\int_{\Omega}Y_{m}^{n}\left(\boldsymbol{k}\right)d\Omega\left(\boldsymbol{k}\right).\label{eq:A7}
\end{equation}
Using the fact that 
\begin{equation}
Y_{m}^{n}\left(\frac{\boldsymbol{n}_{k_{s}}-\boldsymbol{n}_{l_{s}}}{\left\Vert \boldsymbol{n}_{k_{s}}-\boldsymbol{n}_{l_{s}}\right\Vert }\right)=\frac{1}{\sqrt{4\pi}}\quad{\rm {if}}~n=m=0,
\end{equation}
and 
\begin{equation}
\int_{\Omega}Y_{m}^{n}\left(\tilde{\boldsymbol{k}}\right)d\Omega\left(\tilde{\boldsymbol{k}}\right)=\begin{cases}
\sqrt{4\pi} & {\rm {if}}~n=m=0,\\
0 & {\rm {otherwise}},
\end{cases}
\end{equation}
(\ref{eq:A5}) can be rewritten as 
\begin{equation}
\left[\boldsymbol{J}_{s}\right]_{k_{s},l_{s}}=j_{0}\left(2\pi\left\Vert \boldsymbol{n}_{k_{s}}-\boldsymbol{n}_{l_{s}}\right\Vert \right),
\end{equation}
which gives (\ref{eq:1}). Note that (\ref{eq:1}) conforms with \cite{Cook,Robert,9300189}
since $j_{0}\left(c\right)=\frac{\sin c}{c}.$ In addition, (\ref{eq:1})
can be reduced to a 1D fluid antenna with 2D scattering environment
by setting $N_{1}^{s}=1$ and $\frac{0}{0}\triangleq0$ and replacing
$j_{0}\left(\cdot\right)$ by $J_{0}\left(\cdot\right)$ where $J_{0}\left(\cdot\right)$
is the Bessel function of the first kind \cite{9264694}.

\section*{Appendix II: Generalization to Other Spatial Correlation Models}

Without loss of generality, let us consider a 1D fluid antenna since
a similar argument can be made for a 2D fluid antenna surface. To
begin with, let us denote $\boldsymbol{J}$ as the $N\times N$ spatial
correlation matrix. Suppose that $W\gg0$, and the spatial correlation
between the $k$-th port and the $l$-th port is $J_{k,l}=f\left(k,l,N\right)$
where the spatial correlation function $f$ satisfies two conditions:
i) $\lim_{N\rightarrow\infty}f\left(k,k\pm1,N\right)=1$ and ii) there
are some $\exists l\neq k$ such that $f\left(k,l,N\right)\neq1$.
The first condition implies that the $(k\pm1)$-th row/column of $\boldsymbol{J}$
can be removed from $\boldsymbol{J}$ since the $(k\pm1)$-th row/column
of $\boldsymbol{J}$ is always identical to the $k$-th row/column
of $\boldsymbol{J}$ in the limit. The second condition implies that
there exist some $l$-th port whose $l$-th row/column must be retained
in $\boldsymbol{J}$ since its spatial correlation is completely distinct
from the $k$-th port. If $f$ satisfies these two conditions, then
there must exist a minimal spacing $c$ between the $k$-th and $l$-th
ports such that their spatial correlation is distinct. Using the fact
that $W$ is finite, it is clear that there are at most $\bar{N}$
non-identical ports since conditions (i) and (ii) hold. In particular,
$\bar{N}$ must be finite because $W\geq\bar{N}c>0$. As a result,
$\boldsymbol{J}$ can be rewritten as a symmetric $\bar{N}\times\bar{N}$
finite size matrix. Let us denote $N^{'}$ as the full rank of the
symmetric $\bar{N}\times\bar{N}$ matrix where $N^{'}\leq\bar{N}$.
Then, we can further reduce the symmetric $\bar{N}\times\bar{N}$
matrix to a full rank symmetric $N^{'}\times N^{'}$ submatrix $\boldsymbol{J}_{{\rm red}}$
by removing the $(\bar{N}-N^{'})$ dependent rows and columns. Consequently,
$\boldsymbol{J}$ can be represented by $\boldsymbol{J}_{{\rm red}}$
which is a full rank symmetric $N^{'}\times N^{'}$ finite-size matrix.

\section*{\textcolor{black}{Appendix III: The Effect of Mutual Coupling}}

\textcolor{black}{In liquid-based fluid antenna, mutual coupling only
occurs between the active ports. Thus, the MIMO-FAS channel with mutual
coupling effect can be modeled as \cite{8058474} 
\begin{equation}
\boldsymbol{\bar{H}}_{mc}=\boldsymbol{Z}_{mc}^{rx,l}\boldsymbol{\bar{H}}\boldsymbol{Z}_{mc}^{tx,l},\label{eq:C1}
\end{equation}
where $\boldsymbol{Z}_{mc}^{rx,l}$ and $\boldsymbol{Z}_{mc}^{tx,l}$
are the mutual coupling matrices which can be pre-computed offline
given the antenna technologies used. The mutual coupling matrix is
given as 
\begin{equation}
\boldsymbol{Z}_{mc}^{s,l}=\left(Z_{A}+Z_{L}\right)\left(\boldsymbol{Z}_{s}^{l}+Z_{L}\boldsymbol{I}\right)^{-1},s\in\left\{ rx,tx\right\} ,\label{eq:C2}
\end{equation}
where $Z_{A}$, $Z_{L}$ and $\boldsymbol{Z}_{s}^{l}$ are the antenna
impedance, load impedance and mutual impedance matrix of the active
ports, respectively. To compute $\boldsymbol{Z}_{s}^{l}$, we assume
that each active port is a dipole element with a length of $0.5\lambda$
and a width of $0.001\lambda$. Consequently, $Z_{A}=73.08+42.21j$
and $Z_{L}=Z_{A}^{*}$. Due to the dipole's physical constraint, we
fix $N_{1}^{s}=\left\lfloor \frac{W_{1}^{s}}{0.5}\right\rfloor +1$
and vary $N_{2}^{s}$ accordingly. If all the active ports are far
from each other, we have $\boldsymbol{Z}_{mc}^{s}\approx\boldsymbol{I}$.
Given (\ref{eq:C1}), SVD and waterfilling power allocation are then
performed. It is worth highlighting that this model is a conservative
one because the mutual coupling effect is not considered when optimizing
$A_{tx}$ and $A_{rx}$. In fact, the performance of MIMO-FAS can
be further improved when considering the mutual coupling effect and
allowing the optimal active ports to be freely located within the
given surface.}

\textcolor{black}{In RF pixel-based fluid antenna, mutual coupling
can exist regardless of whether the pixels are on or off. Thus, in
practice, it is important to improve the S-matrix via antenna design.
To a coarse approximation, the S-matrix is modeled as 
\begin{equation}
\boldsymbol{S}_{mc}^{s}=\left[\begin{array}{cccc}
\alpha_{\text{rl}}S_{1,1}^{s} & \alpha_{\text{iso}}S_{1,2}^{s} & \cdots & \alpha_{\text{iso}}S_{1,N_{s}}^{s}\\
\alpha_{\text{iso}}S_{2,1}^{s} & \alpha_{\text{rl}}S_{2,2}^{s} &  & \vdots\\
\vdots &  & \ddots\\
\alpha_{\text{iso}}S_{N_{s},1}^{s} & \cdots &  & \alpha_{\text{rl}}S_{N_{s},N_{s}}^{s}
\end{array}\right],\label{eq:C3}
\end{equation}
where $\alpha_{\text{rl}}$ and $\alpha_{\text{iso}}$ determine the
improvement level of the return loss and isolation, respectively,
while $S_{k_{s},l_{s}}^{s}$ is the S-parameter between the $k_{s}$-th
and $l_{s}$-th ports. In this paper, we assume that the return loss
and isolation levels are $-15$dB and $30$dB, respectively, which
are typical values that can be achieved using state-of-the-art technologies
\cite{4685877}. Given (\ref{eq:C3}), the mutual impedance matrix
of the RF pixel-based fluid antenna can be computed as 
\begin{equation}
\boldsymbol{Z}_{s}^{p}=z_{0}\left(\boldsymbol{I}+\boldsymbol{S}_{mc}^{s}\right)\left(\boldsymbol{I}-\boldsymbol{S}_{mc}^{s}\right)^{-1},\label{eq:C4}
\end{equation}
where $z_{0}=50\Omega$ is the reference impedance. Similar to (\ref{eq:C2}),
the mutual coupling matrix is given as 
\begin{equation}
\boldsymbol{Z}_{mc}^{s,p}=\left(Z_{A}+Z_{L}\right)\left(\boldsymbol{Z}_{s}^{p}+Z_{L}\boldsymbol{I}\right)^{-1},s\in\left\{ rx,tx\right\} .\label{eq:C5}
\end{equation}
Similar to (\ref{eq:C1}), the MIMO-FAS channel with mutual coupling
effect is modeled as 
\begin{equation}
\boldsymbol{H}_{mc}=\boldsymbol{Z}_{mc}^{rx,p}\boldsymbol{H}\boldsymbol{Z}_{mc}^{tx,p}.\label{eq:C6}
\end{equation}
In contrast to (\ref{eq:C1}) and (\ref{eq:C2}), it is worth noting
that (\ref{eq:C5}) and (\ref{eq:C6}) are $N_{s}\times N_{s}$ matrices.
Given (\ref{eq:C6}), the port selections, beamforming and power allocation
can be performed by different schemes. Note that matching networks
can be employed to further improve the performance of MIMO-FAS \cite{1310320}.}

\vspace{-5mm}
 \bibliographystyle{IEEEtran} %\bibliography{fluid_antenna}

\providecommand{\url}[1]{#1} \csname url@samestyle\endcsname \providecommand{\newblock}{\relax}
\providecommand{\bibinfo}[2]{#2} \providecommand{\BIBentrySTDinterwordspacing}{\spaceskip=0pt\relax}
\providecommand{\BIBentryALTinterwordstretchfactor}{4} \providecommand{\BIBentryALTinterwordspacing}{\spaceskip=\fontdimen2\font plus
\BIBentryALTinterwordstretchfactor\fontdimen3\font minus
  \fontdimen4\font\relax} \providecommand{\BIBforeignlanguage}[2]{{%
\expandafter\ifx\csname l@#1\endcsname\relax
\typeout{** WARNING: IEEEtran.bst: No hyphenation pattern has been}%
\typeout{** loaded for the language `#1'. Using the pattern for}%
\typeout{** the default language instead.}%
\else
\language=\csname l@#1\endcsname
\fi
#2}} \providecommand{\BIBdecl}{\relax} \BIBdecl 

\bibliographystyle{IEEEtran}
\end{document}